\documentclass[12pt]{article}
\usepackage{amsfonts,amsmath,amssymb,epsf}
\usepackage{amsmath}
\usepackage{graphicx,color}
\newcommand{\be}{\begin{equation}}
\newcommand{\ba}{\begin{eqnarray}}
\newcommand{\ea}{\end{eqnarray}}
\newcommand{\ee}{\end{equation}}

\begin{document}

\begin{titlepage}
\thispagestyle{empty}

\begin{flushright}
YITP-14-41
\end{flushright}


\begin{center}
\noindent{\large \textbf{Notes on Quantum Entanglement of Local
Operators}}\\
\vspace{2cm}

Masahiro Nozaki $^{a}$ 
\vspace{1cm}

{\it
 $^{a}$Yukawa Institute for Theoretical Physics,
Kyoto University, Kyoto 606-8502, Japan\\}

\vskip 2em
\end{center}

\begin{abstract}

This is an expanded version of the short report arXiv:1401.0539, where we studied the (Renyi) entanglement entropies for the excited state defined by acting a given local operator on the ground state. We introduced the (Renyi) entanglement entropies of given local operators which measure the degrees of freedom of local operators and characterize them in conformal field theories from the viewpoint of quantum entanglement. 
In present paper, we explain how to compute them in free massless scalar field theories and we also investigate their time evolution. The results are interpreted in terms of relativistic propagation of an entangled pair. 
The main new results which we acquire in the present paper are as follows.
Firstly, we provide an explanation which shows that the (Renyi) entanglement entropies of a specific operator are given by (Renyi) entanglement entropies of binomial distribution by the replica method. That operator is constructed of only scalar field. Secondly, we found the sum rule which (Renyi) entanglement entropies of those local operators obey. Those local operators are located separately.
Moreover we argue that (Renyi) entanglement entropies of specific operators in conformal field theories are given by (Renyi) entanglement entropies of binomial distribution. These specific operators are constructed of single-species operator. We also argue that general operators obey the sum rule which we mentioned above.

\end{abstract}

\end{titlepage}

\newpage
\section{Introduction}
Recently (R$\acute{e}$nyi) entanglement entropy has become a center of wide interest in a broad array of theoretical physics such as string theory, condensed matter physics and quantum information theory. It is useful tool when we investigate the distinctive features of various quantum states. For example, it provides us with the method of classifying the quantum structure of several states in condensed matter physics, e.g., \cite{wen,pre,mul,hal,wen2,sac2,gon2}. 
The quantum entanglement is expected to be an important quantity which may shed light on the mechanism behind the AdS/CFT correspondence, e.g.,\cite{sw1, sw2, r}. In particular, the entanglement on the boundary is expected to be related to gravity in bulk e.g., \cite{t4,t1,t2,t3}. Therefore, it is important to reveal the fundamental properties of (R$\acute{e}$nyi) entanglement entropy.

Its remarkable feature has been found through researches \cite{thm,hw,son,caputa,pand} over the past a few years. 
A state is excited perturbatively so that the energy of its excitation is much smaller than the inverse of the subsystem's size.
If the size of a subsystem is small, its entanglement entropy is proportional to the energy included in it.
Thus its entanglement entropy obeys the law analogous to the first law of thermodynamics. In relativistic setup, its effective temperature is proportional to the inverse of the subsystem's size. And that temperature does not depend on the parameters of strongly coupled gauge theories in large $N$ limit. In other words, it is universal in strongly coupled gauge theories in this limit. 

In those works, we studied the property of the entanglement entropy when the size of the subsystem is very small.
In this paper and previous one \cite{pne}, we study the property of (R$\acute{e}$nyi) entanglement entropy when its size is infinite. Moreover we define the excited state by acting various local operators on the ground state.
(R$\acute{e}$nyi) entanglement entropies for them show nontrivial time evolution.
Their excess finally approach certain finite values.

Before we explain how to compute them and explain the results which we obtain, we would like to explain quantum quenches briefly. 
There are two classes of quenches; global quenches and local quenches.
 The global quenches are triggered by changing parameters homogeneously, e.g., \cite{cag}. When parameters are changed homogeneously, we are able to study the thermalization of a subsystem and its thermalization time by measuring its entanglement entropy. 
Their holographic duals are the formation of a black hole, e.g., \cite{hgq,hm}. 
The local quenches are triggered by adding an interaction locally to a Hamiltonian or changing parameters locally in the Hamiltonian \cite{cal,eis}.
When parameters are changed inhomogeneously, we can investigate the time evolution of the entanglement entropy locally.  A falling particle in AdS spacetime is proposed as the holographic duals of these quenches, e.g., \cite{hw,uga}. 

In this paper, we introduce a new class of excited states. 
These states are defined by acting a local operator on the ground state. This local operator violates the time translational invariance and excites the ground state. In this setup, we can study the time evolution of (R$\acute{e}$nyi) entanglement entropies for these states.
Therefore we can investigate how the entanglement propagates locally.

We would like to briefly explain how to compute the (R$\acute{e}$nyi) entanglement entropies for these states and give an explanation of the results we obtain. 
In this work, we compute them in the free massless scalar field theory by the replica method. Here we choose the subsystem A to be a half of the total space. We define a locally excited state by acting a local operator $:\phi^k:$ on the ground state. We compute the (R$\acute{e}$nyi) entanglement entropies for these states by using the replica method. After computing them, we perform an analytic continuation to real time and study their time evolution. 
We find they finally approach certain finite constants. 
The time evolution of the (R$\acute{e}$nyi) entanglement entropies for locally excited states can be interpreted in terms of the relativistic propagation of entangled pairs.

 In previous paper \cite{pne}, we derived the (R$\acute{e}$nyi) entanglement entropies at late time for the state which is excited by $:\phi^k:$ from the entangled pair interpretation. 
We call the final values of (R$\acute{e}$nyi) entanglement entropies  ({\it R}$\acute{e}${\it nyi}) {\it entanglement entropies of operators}. They include the (R$\acute{e}$nyi) entanglement entropy for an EPR state.
They measure the degrees of freedom of local operators and characterize them in conformal field theories in the viewpoint of quantum entanglement.

Finally we would like to summarize the new results we obtain in the present work.
In previous paper \cite{pne}, we found that the (R$\acute{e}$nyi) entanglement entropies of $:\phi^k:$ is given by those of binomial distribution only when the replica number and $k$ is small. In present paper, we provide the explanation which shows that the (R$\acute{e}$nyi) entanglement entropies of $:\phi^k:$ are given by those of binomial distribution by using the replica method. The results which we obtain by using replica method agree with those which we obtain in terms of the entangled pair.

We also obtain the sum rule for the (R$\acute{e}$nyi) entanglement entropies for the states defined by acting the various operators on the ground state. They are given by the sum of the (R$\acute{e}$nyi) entanglement entropies for the states defined by acting 
one of these operators on the ground state. 

Moreover we argue that they can be generalized.
We limit operators to the operators $(:\left(\partial^m \phi\right)^k:)$ which are composed of single-species operator.
The (R$\acute{e}$nyi) entanglement entropies of them are given by those of binomial distribution. 
The entropies of general operators obey the same sum rule as the one which we mentioned above.

This paper is organized as follows. In section $2$, we compute the (R$\acute{e}$nyi) entanglement entropies for the states defined by acting various local operators on the ground state in the replica method. We also compute the Green function on the space which has a conical singularity. We explain an analytic continuation which is useful to investigate the time evolution of those entropies for excited states.
In section $3$, we calculate (R$\acute{e}$nyi) entanglement entropies for the states which excited by scalar operators and investigate their time evolution explicitly . 
In section $4$, we interpret their time evolution in terms of entangled pair. We find that the (R$\acute{e}$nyi) entanglement entropies of $:\phi^k:$ is given by those of binomial distribution.
In section $5$, we explain the new results we obtain in this paper. We provide the explanation which shows that the (R$\acute{e}$nyi) entanglement entropies of $:\phi^k:$ are given by those of binomial distribution by the replica method. We argue that those results can be generalized.
In section $6$, we summarize the conclusions and discuss the future problems. 

\section{The R$\acute{e}$nyi entanglement Entropies for Locally Excited States and Propagators }
Let us study the time evolution of (R$\acute{e}$nyi) entanglement entropies for the states generated by acting various local operators on the ground state. To investigate their time evolution, we will compute them by the replica methods.
After computing the (R$\acute{e}$nyi) entanglement entropies for those states by the replica method, we will perform an analytic continuation to real time and investigate their real time evolution. In this section, we will explain the analytic continuation which is used to investigate their time evolution and also explain how to compute them. Those entropies are related to the correlation functions of these operators as we explain later.
In this paper, we compute them in free massless scalar field theory in even dimension.
 Therefore we compute the Green function on $n$-sheeted geometry in even dimensional free massless scalar field theory.


\subsection{How to Investigate the Time Evolution of (R$\acute{e}$nyi) Entanglement Entropies}
\subsubsection{An Analytic Continuation to Real Time}



In this subsection, we would like to explain the analytic continuation which is used to investigate the real time evolution of the (R$\acute{e}$nyi) entanglement entropies for the states defined by acting various operators on the ground state. We consider a QFT in $d+1$ dimensional spacetime $(t, x^i)$~$(i= 1, \cdots,d )$. 
Let us define the locally excited state by
\begin{equation}\label{exm}
\left|\Psi_m\right \rangle = \mathcal{N}^{-1} \mathcal{O}(-t_1,-l,{\bf x}_1)\left| 0 \right \rangle,
\end{equation}
where $\mathcal{O}$ is located at $t=-t_1, x^1=-l, {\bf x}={\bf x}_1$~(${\bf x}=(x^2, x^3, \cdots, x^d)$). 
Its density matrix is given by
\begin{equation}
\hat{\rho} =\mathcal{N}^{-2}\mathcal{O}(-t_1, -l, {\bf x}_1)\left|0\right\rangle \left\langle 0 \right| \mathcal{O}^{\dagger}(-t_1 ,-l, {\bf x}_1).
\end{equation}
where $\mathcal{N}$ is determined so as to $tr \hat{\rho}=1$.
This local operator is acted on the ground state at $t=-t_1$ and triggers the time evolution of quantum entanglement in the total system. 
We investigate the time evolution of quantum entanglement by measuring (R$\acute{e}$nyi) entanglement entropy at $t=0$.

To compute the (R$\acute{e}$nyi) entanglement entropy for these locally excited states, we firstly introduced the regularization parameter $\epsilon$.
By using this parameter, $\hat{\rho}$ is deformed as follow,
\begin{equation}
\hat{\rho} =\mathcal{N}^{-2}e^{-i H t_1}e^{-\epsilon H }\mathcal{O}(-l,{\bf x }_1)\left| 0\right \rangle  \left \langle 0\right| \mathcal{O}^{\dagger} (-l,{\bf x }_1) e^{-\epsilon H } e^{i H t_1}.
\end{equation}
High energy modes are suppressed by $e^{-\epsilon H}$. Thus its (R$\acute{e}$nyi) entanglement entropies are regulated.

To compute the (R$\acute{e}$nyi) entanglement entropies for this state, we use the replica methods.
To compute them by the replica method, we define a density matrix by
\begin{equation}\label{rhoe2}
\hat{\rho} =\mathcal{N}^{-2} e^{\tau_e H}\mathcal{O}(-l,{\bf x }_1)\left| 0 \right\rangle \left \langle 0\right| \mathcal{O}^{\dagger} (-l,{\bf x }_1) e^{-\tau_l H }=\mathcal{N}^{-2}\mathcal{O}(\tau_e, -l , {\bf x}_1)\left|0\right\rangle \left \langle 0\right| \mathcal{O}^{\dagger}(\tau_l, -l, {\bf x}_1).
\end{equation}
where $\tau_e = -\epsilon$, $\tau_l = \epsilon$ respectively. 
 $\epsilon$ is the regularization parameter for two point function of $\mathcal{O}$ and $\mathcal{O}^{\dagger}$.

After computing (R$\acute{e}$nyi) entanglement entropies for this state by replica method, we perform the analytic continuation to real time as follows,
\begin{equation}\label{ac}
\begin{split}
& \tau_e =-\epsilon - i t_1,\\
&\tau_l = \epsilon -i t_1. \\ 
\end{split}
\end{equation} 
After analytic continuation, the relation between the polar coordinate and $\epsilon, t_1$ is shown in Fig.1.
Thus we investigate the evolution of $\Delta S^{(n)}_A$ for excited states given by (\ref{exm}) with $t_1$.

\begin{figure}\label{pro2}	
  \centering
  \includegraphics[width=8cm]{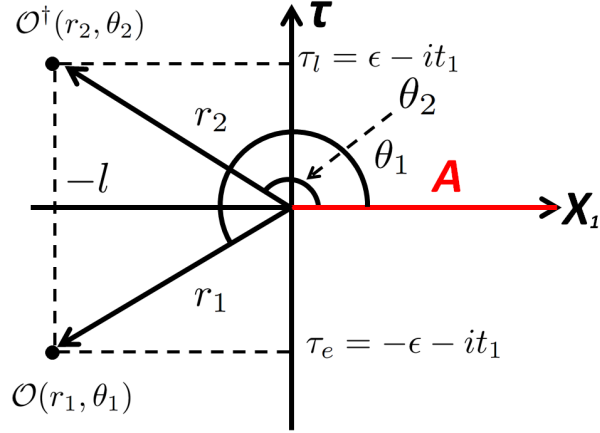}
  \caption{After analytic continuation, $r_1, r_2, \theta_1,$ and $\theta_2$ are related to $\epsilon, t_1$ and $l$ as in this figure.}
\end{figure}

\subsubsection{To Compute the (R$\acute{e}$nyi) Entanglement Entropies By Using Replica Methods}

Let us compute (R$\acute{e}$nyi) entanglement entropies for the density matrix defiend by (\ref{rhoe2}).
The local operators $\mathcal{O}$ and $\mathcal{O}^{\dagger}$ are located at $(\tau = \tau_e,  x^1 =-l,  {\bf x}={\bf x}_1)$ and $(\tau = \tau_l,  x^1 =-l,  {\bf x}={\bf x}_1)$ respectively.

To define (R$\acute{e}$nyi) entanglement entropies, we choose subsystem A to be a half space $x^1\ge 0$\footnote[1]{Of course, we are able to choose any shape as subsystem A. We will compute the (R$\acute{e}$nyi) entanglement entropy for the region $x^1\ge 0$ later. Therefore we choose subsystem A  to be a half space of the total space.}.
Tracing out the degrees of freedom in subsystem B, we define a reduced density matrix by
\begin{equation}
\hat{\rho}_A = tr_B \hat{\rho}.
\end{equation}

The n-th R$\acute{e}$nyi entanglement entropy for this state is defined by
\begin{equation}\label{drenyi}
S^{(n)}_A=\frac{1}{1-n}\log{tr_A \hat{\rho}_A^n}.
\end{equation} 
We would like to study the excess of (R$\acute{e}$nyi) entanglement entropies for this state defined by ($\ref{rhoe2}$). Therefore we subtract those entropies for the ground state from the entropies for the locally excited state and we define their excess  by
\begin{equation}\label{sre}
\Delta S_A^{(n)} = \frac{1}{1-n}\log{\left(\frac{tr_A\hat{\rho}^n_A}{tr_A \hat{\rho}^n_{0A}}\right) }. 
\end{equation}
$\hat{\rho}_{0}$ is the normalized reduced density matrix for the ground state. 

Let us compute $\Delta S^{(n)}_A$ by path-integral as in \cite{RT,rcq,ffac}.
It is useful to introduce a polar coordinate.
The location of operators is shown in Fig.1.

The wave functional $\Psi (\Phi_0(x^i))$ for this state is given by path-integrating from $\tau = -\infty$ to $\tau= 0$ by the replica method.
This wave function $\Psi (\Phi_0(x^i))$ is given by 
\begin{equation}
\Psi(\Phi_0(x^i))= \int^{\Phi(\tau=0, x^i)= \Phi_0(x^i)}_{\Phi(\tau=-\infty, x^i)}D\Phi~ \mathcal{O}(r_1, \theta_1)e^{-S[\Phi]},
\end{equation}
where $\Phi$ is general field of the theory. 
Then the components of the reduced matrix is given by,
\begin{equation}
\begin{split}
&\left[ \hat{\rho}_A \right]_{\Phi_1(x^i)\Phi_2(x^i)} \\
&~~~~=(Z^{EX}_1)^{-1}\int^{\Phi(\infty, x^i)}_{\Phi(-\infty, x^i)}D \Phi~ \mathcal{O}^{\dagger}(r_2, \theta_2,)\mathcal{O}(r_1, \theta_1 )e^{-S \left[ \Phi \right]}\delta\left(\Phi(-\mathcal{\delta}, x^i)-\Phi_1(x^i)\right)\cdot \delta\left(\Phi(\mathcal{\delta}, x^i)-\Phi_2(x^i)\right),
\end{split}
\end{equation}
where we introduce a lattice spacing $\mathcal{\delta}(\ll 1)$.
$Z^{EX}_1$ is defined by
\begin{equation}
Z^{EX}_1=\int^{\Phi(\infty, x^i)}_{\Phi(-\infty, x^i)}D \Phi~ \mathcal{O}^{\dagger}(r_2, \theta_2)\mathcal{O}(r_1,\theta_1)e^{-S(\Phi)},
\end{equation}

Then the $tr \hat{\rho}_A^n$ is given by the partition function where $2n$ operators are inserted, 
\begin{equation}\label{ep}
\begin{split}
&tr_A \hat{\rho}_A^n = \frac{Z^{EX}_n}{\left(Z^{EX}_1\right)^n}, \\
&=\left(Z^{EX}_1\right)^{-n}\int^{\Phi(\infty, x^i)}_{\Phi(-\infty, x^i)}D \Phi~ \mathcal{O}^{\dagger}(r_2, \theta_2)\mathcal{O}(r_1, \theta_1)\cdots \mathcal{O}^{\dagger}(r_2,\theta_{2,k})\mathcal{O}(r_1,\theta_{1,k})\cdots\mathcal{O}^{\dagger}(r_2,\theta_{2,n})\mathcal{O}(r_1,\theta_{1,n}) e^{-S[\Phi]},
\end{split}
\end{equation}
where operators are periodically located on the $n-$sheeted geometry $\Sigma_n$ and $\theta_{i,k}=\theta_i +2\pi (k-1)~ (i=1, 2, k=1 \cdots n)$.
$\Sigma_n$ is constructed gluing the subsystem A on a sheet to subsystem A on the next sheet as in Fig.2.
 \begin{figure}\label{rep}
  \centering
  \includegraphics[width=8cm]{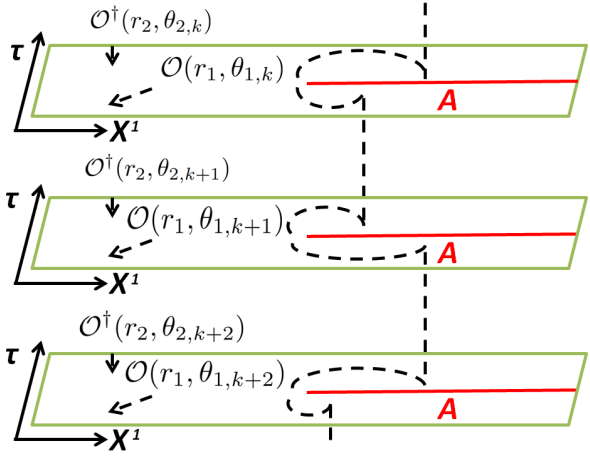}
  \caption{$n$-sheeted geometry $\Sigma_n$ is constructed by gluing subsystem A on a sheet to subsystem A on the next sheet. Fields acquire the phase shift of $2\pi$ if they go around the boundary of A on the each sheet of $\Sigma_n$.}
\end{figure}
And the reduced density matrix for the ground state is given by
\begin{equation}\label{gp}
tr_A \hat{\rho}_{0A}^n = \frac{Z_n}{Z_1}.
\end{equation}
where $Z_n$ is the partition function on the $n-$sheeted geometry and $Z_1$ is the partition function on $R^{d+1}$.

Substituting  (\ref{ep}) and (\ref{gp}) into (\ref{sre}), The path integral representation of $\Delta S_A^{(n)}$ is given by 
\begin{equation}\label{gre}
\begin{split}
&\Delta S_A^{(n)} = \frac{1}{1-n}\left(\log{\frac{tr_A \rho_A^n}{\rho_{0A}}}-n\log{\frac{tr\rho}{tr\rho_0}}\right)=\frac{1}{1-n}\left(\log{\frac{Z^{EX}_n}{Z_n}}-n\log{\frac{Z^{EX}_1}{Z_1}}\right) \\
&=\frac{1}{1-n}\left( \log{\left \langle \mathcal{O}^{\dagger}(r_2, \theta_{2,n})\mathcal{O}(r_1, \theta_{1,n})\cdots \mathcal{O}^{\dagger}(r_2,\theta_{2,1})\mathcal{O}(r_1,\theta_{1,1}) \right\rangle_{\Sigma_n}}-n \log{\left\langle\mathcal{O}^{\dagger}(r_2, \theta_{2,1})\mathcal{O}(r_1,\theta_1) \right\rangle_{\Sigma_1}}\right).
\end{split}
\end{equation}
The first term of the last line in (\ref{gre})  is given by a $2n$ point function of $\mathcal{O}$ and $\mathcal{O}^{\dagger}$ on $\Sigma_n$.
And the second term in (\ref{gre}) is given by $2$ point function of $\mathcal{O}$ and $\mathcal{O}^{\dagger}$ on $\Sigma_1$.
Here $\Sigma_1$ is usual $d+1$ dimensional Euclidean space\footnote[2]{The authors in \cite{Alc,Ast} has found this relation between the R$\acute{e}$nyi entropy and the expectation values of operators.}. 

Here only one operator is acted on the ground state.
However the result in (\ref{gre}) can be generalized to that for the excited state defined by acting several operators on ground state as we will explain later. 

\subsection{Propagators on $\Sigma_n$}

In previous subsection, we defined (R$\acute{e}$nyi) entanglement entropy for locally excited states in general field theories.
From now on, we will explain how to compute those entropies and their evolution with time in free massless scalar field theory.
$\Delta S^{(n)}_A$ are related to the $2n$ point function on $\Sigma_n$ and $2$ point function on $\Sigma_1$ as in (\ref{gre}).
Therefore we explain how to compute the Green function of $\phi$ on even dimensional $\Sigma_n$ \footnote[3]{It is technically difficult to compute the Green function of $\phi$ on odd dimensional $\Sigma_n$. Therefore, we explain how to compute it on even dimensional $\Sigma_n$.}.

We choose the region of subsystem A to be a half $(x^1 \ge 0)$ of the total space as in Fig.3.
We introduce the polar coordinate as $x^1+i\tau =r e^{i\theta}$ $(0\le r<\infty, 0\leq \theta < 2 n\pi)$.

 \begin{figure}\label{pro1}
  \centering
  \includegraphics[width=8cm]{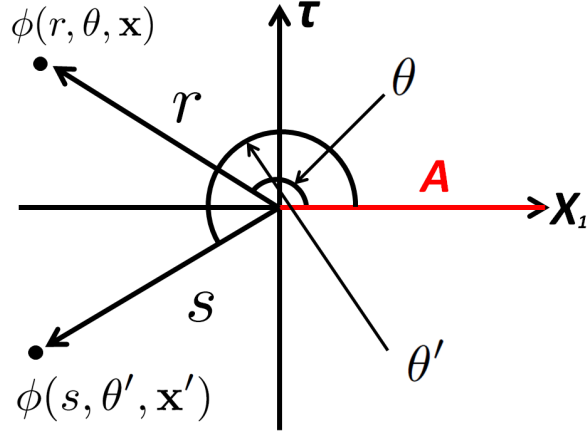}
  \caption{The subsystem A is given by the region $x_1 \ge 0, \tau=0$. Operators are located at $(r, \theta, {\bf x}), (s, \theta' , {\bf x}')$ respectively. ${\bf x}$ are the coordinates along the direction vertical to the plane which has a conical singularity.}
\end{figure}

The Green function on $\Sigma_n$ is defined by 
\begin{equation}
\mathcal{L} G(r, s, \theta, \theta', {\bf x}, {\bf x'})= \left(\partial_r^2+\frac{1}{r}\partial_r +\frac{1}{r^2}\partial_{\theta}^2+\partial_{\bf x}^2\right) G(r, s, \theta_a, \theta_b, {\bf x}, {\bf x'})= -\delta(x-x'),
\end{equation}
where ${\bf x}= (x^2, x^3, \cdots ,x^d)$ and $x= (\tau, x^1, x^2, x^3, \cdots ,x^d)$.
Then we can expand the Green function in terms of the eingenfunctions $v(r, \theta, {\bf x}) $, 
\begin{equation}
\mathcal{L}v(r, \theta, {\bf x})=\lambda v(r, \theta, {\bf x}),
\end{equation}
where $v(r, \theta, {\bf x})=v(r, \theta+2n\pi, {\bf x})$.
As in \cite{capro}, the Green function on $\Sigma_n$ is given by
\begin{equation}\label{1}
G(r, s, \theta, \theta', {\bf x}, {\bf x'})=\frac{1}{2\pi n}\sum^{\infty}_{l=0}d_l \int ^{\infty}_0dk\int \frac{dk^{d-1}_{\perp}}{(2\pi)^{d-1}}\frac{k\cdot J_{\frac{l}{n}}(k r)J_{\frac{l}{n}}(k s)}{k^2+k_{\perp}^2}e^{i{\bf k}_{\perp}\cdot \left({\bf x}'-{\bf x}\right)}\cos{\left(\frac{\theta-\theta'}{n}l\right)}.
\end{equation}
where $d_0 =1, d_{l>0}=2$ and $J_{\nu}(x)$ is the Bessel function of the first kind.

To integrate with respect to $k_{\perp}$ in right hand side of the equation in (\ref{1}), we express it in terms of a Schwinger parameter.
After that, its denominator is given by
\begin{equation}
\frac{1}{k^2+k_{\perp}^2}=\int ^{\infty}_{0}du e^{-(k^2+k_{\perp}^2)u}.
\end{equation}

After express it in terms of Schwinger parameter, we integrate with respect to $k_{\perp}$ in the right hand side of that equation,
\begin{equation}
\int \frac{dk_{\perp}^{d-1}}{(2\pi)^{d-1}} e^{-u(k_{\perp})^2+i{\bf k}_{\perp}\cdot \left({\bf x}'-{\bf x}\right)}=\frac{1}{(2\sqrt{u \pi})^{d-1}}e^{-\frac{1}{4u}\left({\bf x}'-{\bf x}\right)^2}.
\end{equation}

Moreover, we integrate with respect to $k$.
If we use the formula of Bessel function, the integral of $k$ from $0$ to $\infty$ in (\ref{1}) can be performed,  
\begin{equation}
\int^{\infty}_{0}dk k e^{-u k^2}J_{\frac{l}{n}}(k r)J_{\frac{l}{n}}(k s)=\frac{1}{2u}e^{-\frac{r^2+s^2}{4u}}I_{\frac{l}{n}}\left(\frac{r s}{2u}\right)
\end{equation}
where $I_{\frac{l}{n}}(x)$ is the modified Bessel function of first kind.
$I_{\frac{l}{n}}(x)$ is given by
\begin{equation}
I_{\frac{l}{n}}\left(\frac{r s}{2u}\right)=\frac{1}{2\pi i}\int^{\infty+i\pi}_{\infty-i\pi}dte^{\frac{r s}{2u}\cosh{t}-\frac{l}{n}t},
\end{equation} 
where the contour in the complex plane is shown in Fig.4.

\begin{figure}\label{ccp}	
  \centering
  \includegraphics[width=8cm]{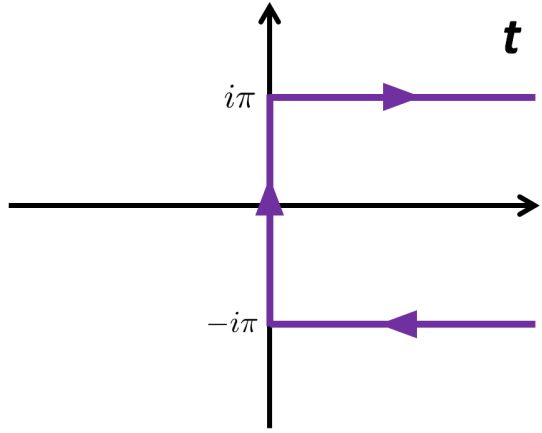}
  \caption{The purple line corresponds to the integral contour in the complex plane.}
\end{figure}

After performing integral of $u$, $G(r, s, \theta, \theta', {\bf x}, {\bf x'})$ is given by
\begin{equation}
\begin{split}
&G(r, s, \theta, \theta', {\bf x}, {\bf x'}) =\frac{\Gamma \left(\frac{d-1}{2}\right)}{4n\pi\left(2\pi^{\frac{1}{2}}\right)^{d-1}} \\
&~~~~\times\frac{1}{2i\pi}\sum_{l=0}^{\infty}d_l \int^{\infty+i\pi}_{\infty-i\pi}dt\frac{4^{\frac{d-1}{2}}e^{-\frac{l}{n}t}\cdot \cos{\left(\frac{l\left(\theta-\theta'\right)}{n}\right)}}{\left\{\left(\left({\bf x}'-{\bf x}\right)^2+r^2+s^2-2r s\cosh{t} \right)\right\}^{\frac{d-1}{2}}}.
\end{split}
\end{equation}
When d is odd, we can perform integration of t. 

Below we will mainly investigate the time evolution of (R$\acute{e}$nyi) entanglement entropy for excited states in $4$ and $6$ dimensional cases. 
To evaluate (\ref{gre}), we would like to give the explicit form of the Green function on $4$ and $6$ dimensional $\Sigma_n$. 
When $d$ is $3$, $G(r, s, \theta, \theta', {\bf x}, {\bf x}')$ is given by
\begin{equation}
G(r, s, \theta, \theta', {\bf x}, {\bf x}') =\frac{1}{4n\pi^2 r s (a-a^{-1})}\frac{a^{\frac{1}{n}}-a^{-\frac{1}{n}}}{a^{\frac{1}{n}}+a^{-\frac{1}{n}}-2\cos{ \left( \frac{\theta-\theta'}{n} \right)}}.
\end{equation}
where the parameter $a$ is given by
\begin{equation}
\begin{split}
\frac{a}{1+a^2}=\frac{r s}{\left|{\bf x'}-{\bf x}\right|^2+r^2+s^2},\\
\end{split}
\end{equation}
as in \cite{sac,jac,lin,dow}. 
When $d$ is $5$, $G(r, s, \theta, \theta', {\bf x}, {\bf x'})$ is given by 
\begin{equation}
\begin{split}
&G(r, s, \theta, \theta', {\bf x}, {\bf x'}) \\
&~~=\frac{1}{4\pi ^3 n^2 (r s)^2\left( a -a^{-1}\right)^2} \\
&\times \left[\frac{2\left((a^{\frac{1}{n}}+a^{-\frac{1}{n}})\cos{\left(\frac{\theta-\theta'}{n}\right)}-2\right)}{\left(a^{\frac{1}{n}}+a^{-\frac{1}{n}}-2\cos{\left(\frac{\theta-\theta'}{n}\right)}\right)^2} +\frac{n(a+a^{-1})(a^{\frac{1}{n}}-a^{-\frac{1}{n}})}{\left(a^{\frac{1}{n}}-a^{-\frac{1}{n}}-2\cos{ \left(\frac{\theta-\theta'}{n}\right)} \right) \left( a-a^{-1} \right) }\right].
\end{split}
\end{equation}
By using these Green functions, we will study the time evolution of (R$\acute{e}$nyi) entanglement entropies for locally excited states.

\section{The Results of R$\acute{e}$nyi Entanglement Entropies}

Let us study the time evolution of (R$\acute{e}$nyi) entanglement entropies for various locally excited states in $2, 4$ and $6$ dimensional free massless scalar field theory.
They will approach a finite constant value at late time.
Here, we will explain how to compute them. 

First, we will study them in two dimensional examples. In these examples, we can compute the R$\acute{e}$nyi entanglement entropies for locally excited states by using a conformal map.
We will explain the results of $\Delta S^{(2)}_A$ for the states defined by acting  several  operators on the ground state. We choose $\partial \phi$, $:\partial \phi \partial \phi:$, $:\bar{\partial}\phi \partial \phi$, $:e^{\pm i \alpha \phi}:$ and $:e^{i \alpha \phi}:+c :e^{-i \alpha \phi}:$ as the operators ($c$ is a complex number). We define the excited state by acting them on the ground state. $\Delta S^{(2)}_A$ for the last one shows nontrivial time evolution. We assume these operators are located at $(t=-t_1, x^1=-l)$.

In 4 and 6 dimensions, we define the excited states by acting $:\phi^k:$ on the ground state. $\Delta S^{(n)}_A$ for those states show the similar time evolution and approach the finite constants at late time in both dimensions. $\Delta S^{(n)}_A$ in $6$ dimension increase milder than those in $4$ dimension. However the final values of $\Delta S^{(n)}_A$ do not depend on spacetime dimension but depend on $k$ and $n$. Therefore we explain would like to explain how to compute them in only $4$ dimension. 

Here we compute them in two illustrative examples.
In the first example, we compute $\Delta S^{(2)}_A$ for the state which is excited by a the local operator $\phi$. We also compute that for the state which is excited by the two local operators $\phi \phi$ in the second example. In the latter example, we assume the two operators are located separately. In the latter example, we would like to study the time evolution of $\Delta S^{(2)}_A$ in the two setups. 

In first setup, one operator $\phi$ is located at $(t=-t_1, x^1=-L)$ and another one is located at $(t=-t_2, x^1=-l)$ where $(l \neq L, t_1 \ge L)$.  In the region $t_2 \le l$, $\Delta S^{(2)}_A$ is given by a finite constant which depends on the parameters $t_1, L$. In the region $t_2 \ge l$,  $\Delta S^{(2)}_A$ shows nontrivial time evolution and finally approaches a certain constant value. It also depends on only two parameters $t_1, L$. In both cases, $\Delta S^{(2)}_A$ is given by the sum of the (R$\acute{e}$nyi) entanglement entropy for the state which is excited by $\phi$.\footnote[4]{ In every cases, we assume $l\ge 0$ and $L\ge 0$ for simplicity.
}


\subsection{The Results in $2$-dimension}

In this subsection, we explain how to compute $\Delta S^{(n)}_A$ for locally excited states and also explain the results in two dimensional examples.
We compute $\Delta S^{(n)}_A$ for the states defined by acting the derivatives of $\phi$ and exponentials of $\phi$ on the ground state.
We choose $\partial \phi$, $:\partial \phi  \partial \phi:$ and $:\bar{\partial} \phi  \partial \phi:$ as derivatives of $\phi$. $:e^{\pm i \alpha \phi}:$ and $:e^{i \alpha \phi}:+c :e^{-i \alpha \phi}:$ are chosen as its exponentials\footnote[5]{In two and higher dimensions, we will study the time evolution of $\Delta S^{(n)}_A$ for the states defined by acting local operators whose conformal dimension is positive on the ground state. 
In two dimension the conformal dimension of $\phi$ vanishes. It does not seem to be a local operator from viewpoint of quantum entanglement. Therefore we will not explain $\Delta S^{(n)}_A$ for the state which excited by $\phi$ in this paper.}.

\subsubsection{How to compute $\Delta S^{(n)}_A$}
In two dimensional free massless scalar filed theory, we can calculate $\Delta S^{(n)}_A$ by using a conformal map $\omega=z^n$.
In this case, we do not need the Green function which we obtain in section 2.
Here $z$ is the coordinate of $n$-sheeted geometry and $\omega$ is the coordinate of the flat plane. To compute derivatives and exponentials of $\phi$, we just have to obtain the two point function of $\phi$. The green function on the flat plane $(=\Sigma_1)$ is given by
\begin{equation}
\left\langle \phi(\omega_1, \bar{\omega}_1)\phi(\omega_2, \bar{\omega}_2)\right\rangle =-\frac{1}{2}\log{\left|\omega_1-\omega_2\right|^2}.
\end{equation}
We impose that the behavior of the two point function on $\Sigma_n$ is the same as the behavior of the Green function on $\Sigma_1$ if we take the $z_1 \rightarrow z_2$ limit. 
We also impose that this two point function on $\Sigma_n$ satisfies the properties of Green function. By using the conformal map $\omega =z^n$, two point function of $\phi$ on $\Sigma_n$ is given by
\begin{equation}\label{epro}
\left \langle \phi  \left( z_1, \bar{z_1} \right)  \phi\left (z_2, \bar{z}_2\right)\right \rangle_{\Sigma_n} =-\frac{1}{2}\log{\left| z_1^{\frac{1}{n}} -z_2^{\frac{1}{n}}\right|}+\frac{1}{2}\log{\left(\frac{\left|z_1\right|^{\frac{1}{n}-1}}{n}\right)} +\frac{1}{2}\log{\left(\frac{\left|z_2 \right|^{\frac{1}{n}-1}}{n}\right)}.
\end{equation}

After substituting the propagators in (\ref{epro}) into (\ref{gre}), we perform the analytic continuation defined in (\ref{ac}) to real time. 
After that, we can study time evolution of $\Delta S_A$ for the states defined by acting local operators on the ground state.

\subsubsection{The Derivatives of $\phi$}

Here we compute the $\Delta S^{(2)}_A$ for locally excited states defined by acting the derivatives of $\phi$ on the ground state.
These operators are limited to $\partial \phi(z,\bar{z})$, $:\partial \phi(z,\bar{z}) \partial \phi(z, \bar{z}):$ , $:\bar{\partial} \phi(z, \bar{z}) \partial \phi(z, \bar{z}):$.

After computing the second R$\acute{e}$nyi entanglement entropies for these states by the replica method,
we perform the analytic continuation in (\ref{ac}) and take the $\epsilon \rightarrow 0$ limit.
In this limit, they vanish,
\begin{equation}
\Delta S^{(2)}_A = 0.
\end{equation}
As we explain in the subsection 4.2, these results can be physically interpreted in terms of relativistic propagation of quasi-particles.

\subsubsection{Exponential operators of $\phi$}
We would like to explain the results of the excited states generated by acting  exponential operators of $\phi$ on the ground state such as $:e^{\pm i \alpha \phi}:$, $:e^{i \alpha \phi }:+c :e^{- i \alpha \phi}:$.
$\Delta S^{(2)}_A$ for the state which is excited by $:e^{\pm i \phi }:$ vanishes.

On the other hand, $\Delta S^{(2)}_A$ for the state defined by acting $:e^{i \alpha \phi }:+c :e^{- i \alpha \phi}:$  on the ground state does not vanish.
In the replica method, $\Delta S^{(2)}_A$ for this state is given by
\begin{equation}\label{2se}
\Delta S^{(2)}_A =-\log{\left[\frac{\left\langle\mathcal{O}^{\dagger}(r_2, \theta_2 +2\pi)\mathcal{O}(r_1, \theta_1+2\pi) \mathcal{O}^{\dagger}(r_2,\theta_2)\mathcal{O}(r_1, \theta_1) \right \rangle_{\Sigma_2}}{\left \langle \mathcal{O}^{\dagger}(r_2, \theta_2) \mathcal{O}(r_1, \theta_1) \right \rangle_{\Sigma_1}}\right]},
\end{equation}
where $\mathcal{O}$ and $\mathcal{O}^{\dagger}$ are given by
\begin{equation}
\begin{split}
&\mathcal{O}(r_1, \theta_1) =:e^{i \alpha \phi }: +c e^{-i \alpha \phi}, \\
&\mathcal{O}^{\dagger}(r_1, \theta_1) =:e^{-i \alpha \phi }: +c^{*} e^{i \alpha \phi}. \\
\end{split}
\end{equation}

By substituting propagators in (\ref{epro}) into (\ref{2se}), 
we perform the analytic continuation which is defined in (\ref{ac}) and take the $\epsilon \rightarrow 0$ limit.
The time evolution of $\Delta S_A$ is given by the step function,
\begin{equation}
\Delta S^{(2)}_A =
\begin{cases}
0 &t_1 \le 0, \\
-\log{\left[\frac{1+\left|c\right|^4}{\left(1+\left|c\right|^2\right)^2}\right]} & t_1 >0. \\
\end{cases}
\end{equation} 
$\Delta S^{(2)}_A$ does not depend on the conformal dimension of operator $\mathcal{O}$.
This final value depends on only the complex constant $c$.
When $\left| c \right|$ is $1$, the final value of $\Delta S^{(2)}_A$ is given by its maximum value .
It is equivalent to the second R$\acute{e}$nyi entanglement entropy for an EPR state.
The time evolution of $\Delta S^{(2)}_A$ is plotted in Fig.5.
In subsection 4.1, we interpret this result in terms of entangled quanta.
\begin{figure}\label{2dex}	
  \centering
  \includegraphics[width=8cm]{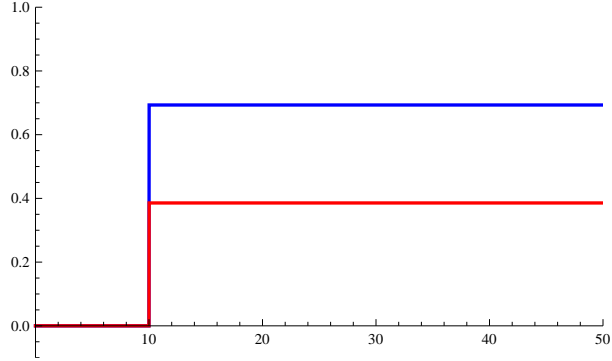}
  \caption{ The plots of $\Delta S^{(2)}_A$ as functions of $t_1$ in the $\epsilon \rightarrow 0$ limit.
The vertical line corresponds to $\Delta S^{(2)}_A$. And the horizontal line corresponds to $t_1$. 
Here we chose $l=10$. The red curve corresponds to the plot of $\Delta S^{(2)}_A$ for $|c| = \frac{1}{2}$.
The blue curve corresponds to the plot of $\Delta S^{(2)}_A$ for $|c| = 1$.}
\end{figure}

\subsection{Higher dimensional cases}
In this subsection, we compute $\Delta S^{(2)}_A$ and study its time evolution in two examples. In both cases, we compute $\Delta S^{(2)}_A$ for locally excited states by the replica method. After computing it, we perform the analytic continuation defined by (\ref{ac}) to real time and we study its real time evolution. 
 
\subsubsection{Single Operator Case}
As we mentioned above, we compute $\Delta S^{(2)}_A$ for the excited state defined by acting a local oprerator $\phi$ in $4$ dimensional free massless scalar field theory.
Let us calculate its second R$\acute{e}$nyi entanglement entropy and study its time evolution. We assume that $\phi$ is located at $\left(t=-t_1,x^1=-l\right)$. This locally excited state is given by 
\begin{equation}
\left|\Psi\right\rangle =\mathcal{N}^{-1} \phi(-t_1,-l)\left|0\right\rangle.
\end{equation}

To investigate the time evolution of $\Delta S^{(2)}_A$ for this locally excited state, we computed $\Delta S^{(2)}_A$ by the replica method.
This operator is located at $\tau = \tau_e,  x^1=-l$. This state is given by 
\begin{equation} \label{lp1}
\left|\Psi\right\rangle =\mathcal{N}^{-1} \phi(\tau_e,-l)\left|0\right\rangle.
\end{equation}

After mapping $(\tau, x^1)$ to $(r, \theta)$ as in Fig.6, $\Delta S^{(2)}_A$ is given by
\begin{equation}\label{phi1}
\Delta S^{(2)}_A = -\log{\left[\frac{\left\langle\phi(r_1,\theta_1)\phi(r_2, \theta_2) \phi(r_1,\theta_1+2\pi)\phi(r_2, \theta_2+2\pi)\right \rangle_{\Sigma_2}}{\left \langle \phi(r_1,\theta_1)\phi(r_2,\theta_2)\right \rangle_{\Sigma_1}^2}\right]},
\end{equation} 
where $\phi(r_2,\theta_2)$ is located at $\left(\tau=\tau_l, x^1=-l\right)$ as in Fig.6.

\begin{figure}\label{smap}	
  \centering
  \includegraphics[width=8cm]{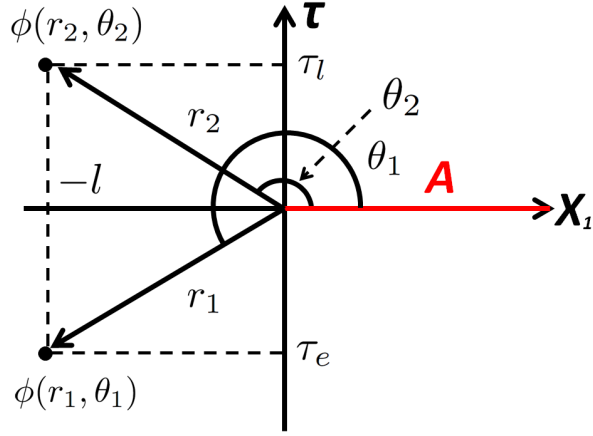}
  \caption{The locations of operators.}
\end{figure}

To compute $\Delta S^{(2)}_A$, we need to compute $4$ point correlation function of $\phi$ on $\Sigma_2$ and $2$ point correlation function of $\phi$ on $\Sigma_1$.
Therefore we compute the Green function on $\Sigma_2$ and $\Sigma_1$. And these Green functions are respectively given by
\begin{equation}\label{4pro}
\begin{split}
\left\langle \phi(r, \theta) \phi(s, \theta' ) \right \rangle_{\Sigma_2} 
&=\frac{1}{8 \pi ^2 (r+s) \left( r+s-2 \sqrt{r s} \cos{\left(\frac{\theta - \theta' }{2}\right)}\right)},\\
\left\langle \phi(r, \theta) \phi(s, \theta' ) \right \rangle_{\Sigma_1} &=\frac{1}{4 \pi ^2 \left(r^2+s^2-2 r s \cos{\left(\theta-\theta'\right)}\right)}.
\end{split}
\end{equation}

By substituting these propagators in (\ref{4pro}) into (\ref{phi1}), we perform the analytic continuation defined by (\ref{ac}) to real time.

In this analytic continuation, the following useful relations hold
\begin{equation}
\begin{split}
&r_1^2 =l^2 +\epsilon^2 -t_1^2 +2 i \epsilon t_1, \\
&r_2^2 =l^2 +\epsilon^2 -t_1^2-2 i \epsilon t_1, \\
&r_1 r_2 \cos{(\theta_1-\theta_2)}=l^2-\epsilon^2 -t_1^2.
\end{split}
\end{equation}

After performing this analytic continuation, we take the $\epsilon \rightarrow 0$ limit. In this limit, the two point function on $\Sigma_{1}$ is given by
\begin{equation}
\left \langle \phi(r_1, \theta_1) \phi(r_2, \theta_2) \right \rangle_{\Sigma_1}=\frac{1}{16\pi ^2\epsilon ^2}.
\end{equation}
And the leading term of four point function on $\Sigma_2$ is $\mathcal{O}(\epsilon^{-4})$ in this limit. Only specific propagators can contribute to this four point function. We call these propagators {\it dominant propagators}. In the region $t_1<l$, dominant propagators are given by
\begin {equation}
\left\langle \phi(r_1, \theta_1)\phi(r_2, \theta_2)\right \rangle_{\Sigma_2 } =\left\langle \phi(r_1, \theta_1+2\pi)\phi(r_2, \theta_2+2\pi)\right \rangle_{\Sigma_2 } =\frac{1}{16 \pi ^2 \epsilon ^2}+\mathcal{O}(\epsilon^0).
\end{equation} 
In the region $t_1\ge l$, dominant propagators are given
\begin {equation}
\begin{split}
\left\langle \phi(r_1, \theta_1)\phi(r_2, \theta_2)\right \rangle_{\Sigma_2} =\left\langle \phi(r_1, \theta_1+2\pi)\phi(r_2, \theta_2+2\pi)\right \rangle_{\Sigma_2} =\frac{l+t_1}{32 \pi ^2 t_1 \epsilon ^2}+\mathcal{O}\left(\epsilon^{0}\right), \\
\left\langle \phi(r_1, \theta_1+2\pi)\phi(r_2, \theta_2)\right \rangle_{\Sigma_2 } =\left\langle \phi(r_1, \theta_1)\phi(r_2, \theta_2+2\pi)\right \rangle_{\Sigma_2 } =\frac{-l+t_1}{32 \pi ^2 t_1 \epsilon ^2}+\mathcal{O}\left(\epsilon^{0}\right). \\
\end{split}
\end{equation} 
The other propagators are at most $\mathcal{O}(\epsilon^0)$ in both regions. Therefore we can ignore them in the $\epsilon \rightarrow 0$ limit.

In the region $t_1 < l$, the four point correlation function of $\phi$ on $\Sigma_2$ is given by
\begin{equation}
\left\langle \phi (r_1,\theta_1) \phi (r_2, \theta_2) \phi (r_1,\theta_1+2\pi) \phi (r_2, \theta+2\pi) \right \rangle_{\Sigma_2} =\frac{1}{256 \pi ^4 \epsilon ^4}+\mathcal{O}(\epsilon^{-3}).
\end{equation}
In the region $t_1\ge l$, that is given by 
\begin{equation}
\left\langle \phi (r_1,\theta_1) \phi (r_2, \theta_2) \phi (r_1,\theta_1+2\pi) \phi (r_2, \theta_2+\pi) \right \rangle_{\Sigma_2} =\frac{l^2+t_1^2}{512 \pi ^4 t_1^2 \epsilon ^4}+\mathcal{O}(\epsilon^{-3}).
\end{equation} 

Therefore, the second R$\acute{e}$nyi entanglement entropy for the excited state defined by acting $\phi$ on the ground state is given by 
\begin{equation}\label{sep1}
\Delta S^{(2)}_A =\log{\left[\frac{2 t_1^2}{l^2+t_1^2}\right]}.
\end{equation}
The time evolution of $\Delta S^{(2)}_A$ in (\ref{sep1}) is plotted in Fig.7.

\begin{figure}\label{sl10}	
  \centering
  \includegraphics[width=8cm]{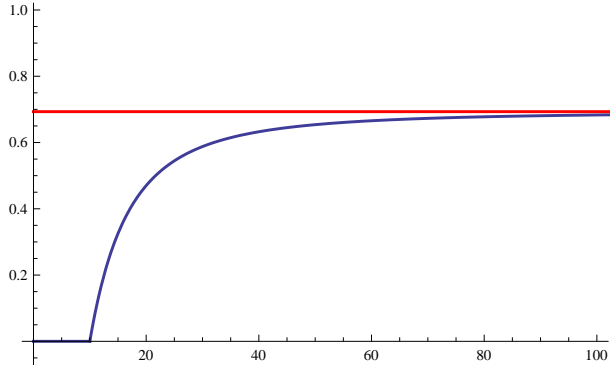}
  \caption{The evolution of $\Delta S^{(2)}_A$ with $t_1$ in the $\epsilon \rightarrow 0$ limit.
The vertical axis corresponds to $\Delta S^{(2)}_A$. The horizontal axis corresponds to $t_1$. Here we chose $l=10$.
The blue curve is the $\Delta S^{(2)}_A$ for $\phi\left|0\right\rangle$. The red curve is the final value of $\Delta S^{(2)}_A$.
It is given by $\log{2}$.
}
\end{figure}

If we take the $t_1\rightarrow \infty$ limit, its final value is given by 
\begin{equation}
\Delta S^{(2) f}_A = \log{2}.
\end{equation}

\subsection{Multiple Operators Case}
Let us compute the second R$\acute{e}$nyi entanglement entropy for the state defined by acting two local operators $\phi \phi$ on the ground state.
We assume these operators are located separately. We compute it in two different examples. In first example, we act $\phi$ on the ground state at time $t= -t_1$ and act another one on it at time $t=-t_2$. We investigate evolution of the second R$\acute{e}$nyi entanglement entropy with time $t=t_2$. In second example, we act two operators on the ground state simultaneously. We also investigate the time evolution of the second R$\acute{e}$nyi entanglement entropy. 

\subsubsection{Not Simultaneous Case}
Let us compute $\Delta S^{(2)}_A$ for the state which is excited by two local operators $\phi \phi$. One of those operators is located at $t=-t_1, x^1 =- L$ as in Fig.8. Here, we assume that $t_1\geq L$. If $t_1 \leq L$, then the time evolution of $\Delta S^{(2)}_A$ is not affected by local operator $\phi(-t_1,-L)$. Another one is located at $t=-t_2, x^1=-l$. We would like to investigate the evolution of $\Delta S^{(2)}_A$ with time $t_2$. 
\begin{figure}\label{mana}	
  \centering
  \includegraphics[width=8cm]{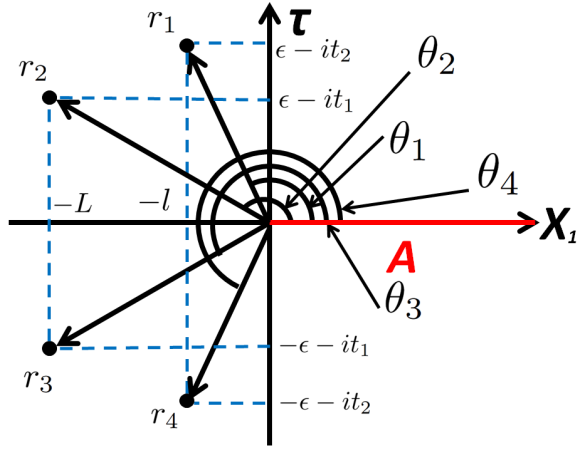}
  \caption{The relation between coordinates. $r, \theta$ is related to $l, L, t_1, t_2$ and $\epsilon$ after performing an analytic continuation to real time.}
\end{figure}

In this example, the excited state is given by 
\begin{equation}
\left|\Psi\right\rangle =\mathcal{N}^{-1}\mathcal{T}\phi(-t_1, -L)\phi(-t_2, -l)\left|0\right\rangle,
\end{equation}
where $\mathcal{T}$ is the time ordering operator.
In the replica method, $\Delta S^{(2)}_A$ for this state is given by
\begin{equation}\label{dtwo}
\begin{split}
&\Delta S^{(2)}_A  \\
&=-\log{ \left[\frac{\left \langle \phi(r_1, \theta_1) \phi(r_2, \theta_2)\phi(r_3,  \theta_3)\phi(r_4,\theta_4) \phi(r_1, \theta_1+2\pi) \phi(r_2, \theta_2+2\pi)\phi(r_3,  \theta_3+2\pi)\phi(r_4,\theta_4+2\pi)\right\rangle_{\Sigma_2}}{\left\langle \phi(r_1, \theta_1) \phi(r_2, \theta_2)\phi(r_3,  \theta_3)\phi(r_4,\theta_4)\right \rangle_{\Sigma_1}^2}\right]}
\end{split}
\end{equation}

To compute $\Delta S^{(2)}_A$, we have to compute the eight point correlation function of $\phi$ on $\Sigma_2$ and the four point function on $\Sigma_1$. 
Therefore we need to compute Green function on $\Sigma_2$ and $\Sigma_1$. These Green functions are given by (\ref{4pro}). 

In the replica method, $\phi(-t_1, -L)$ corresponds to $\phi (r_2, \theta_2)$ and $\phi (r_3, \theta_3)$ and they are located at $(\tau=\tau_{l, 1}, x^1=-L)$ and $(\tau=\tau_{e, 1}, x^1=-L)$ respectively.
On the other hand, $\phi(-t_2, -l)$ corresponds to $\phi(r_1, \theta_1)$, $\phi(r_4, \theta_4)$ in that method and they are located at $(\tau=\tau_{l, 2}, x^1=-l)$ and $(\tau=\tau_{e, 2}, x^1 =-l)$ respectively as in Fig.8.
In addition to the correlations in the case where one operator is acted on the ground state, there are the correlations which correspond to the correlation between $\phi(-t_1, -l)$ and $\phi(-t_2, -L)$ such as $\left \langle \phi(r_1, \theta_1)\phi(r_2, \theta_2)\right\rangle$,  $\left \langle \phi(r_1, \theta_1)\phi(r_2, \theta_2+2\pi)\right\rangle$ $\cdots$.
It seems that we have to take these correlation functions into account. However these correlation functions can not contribute to $\Delta S^{(2)}_A$ when we take the $\epsilon \rightarrow 0$ limit. Therefore we can neglect the contribution from them.
Below we will explain why these correlation functions can not contribute to $\Delta S^{(2)}_A$ in this limit.

After substituting the propagators in (\ref{4pro}) into (\ref{dtwo}), we perform an analytic continuation as follow, 
\begin{equation}\label{2ana}
\begin{split}
&\tau_{l, 1}=\epsilon -i t_{1}, \\
&\tau_{l, 2}=\epsilon -i t_{2}, \\
&\tau_{e, 1}=-\epsilon -i t_{1}, \\
&\tau_{e, 2}=-\epsilon -i t_{2}.\\
\end{split}
\end{equation}
as in Fig.8.
In this continuation, there are useful relations as follows,
\begin{equation}
\begin{split}
&r_1^2 =l^2 +\epsilon^2 -t_2^2-2i \epsilon t_2, \\
&r_2^2 =L^2 +\epsilon^2 -t_1^2-2i \epsilon t_1, \\
&r_3^2 =L^2 +\epsilon^2 -t_1^2+2i \epsilon t_1, \\
&r_2^2 =l^2 +\epsilon^2 -t_2^2+2i \epsilon t_2, \\
\end{split}
\end{equation}
and
\begin{equation}
\begin{split}
&r_1 r_2 \cos{\left(\theta_1-\theta_2\right)}=l L+\epsilon^2-t_1 t_2 -i\epsilon (t_1+t_2), \\
&r_1 r_3 \cos{\left(\theta_1-\theta_3\right)}=l L-\epsilon^2-t_1 t_2 -i\epsilon (t_1-t_2), \\
&r_1 r_4 \cos{\left(\theta_1-\theta_4\right)}=l^2-\epsilon^2-t_2^2 , \\
& r_2 r_3 \cos{\left(\theta_2-\theta_3\right)}=L^2-\epsilon^2-t_1^2 , \\
&r_2 r_4 \cos{\left(\theta_2-\theta_4\right)}=l L-\epsilon^2-t_1 t_2 -i\epsilon (-t_1+t_2), \\
&r_3 r_4 \cos{\left(\theta_3-\theta_4\right)}=l L+\epsilon^2-t_1 t_2 +i\epsilon (t_1+t_2). \\
\end{split}
\end{equation}
After performing this analytic continuation, we take the limit $\epsilon \rightarrow 0$. In this limit, 4 point function on $\Sigma_1$ is given by 
\begin{equation}
\left\langle \phi(r_1, \theta_1) \phi(r_2, \theta_2)\phi(r_3,  \theta_3)\phi(r_4,\theta_4)\right \rangle_{\Sigma_1} =\frac{1}{256 \pi ^4 \epsilon ^4}+\mathcal{O}\left(\epsilon^0\right ).
\end{equation}

On the other hand, the leading term of 8 point function of $\phi$ on $\Sigma_2$ is $\mathcal{O}(\epsilon^{-8})$ in this limit. In the region $t_2 \leq l$, the leading term of 8-point function of $\phi$ is given by 
\begin{equation}
\left \langle \phi^8 \right \rangle_{\Sigma_2} \sim
\frac{L^2+t_1^2}{131072 \pi ^8 t_1^2 \epsilon ^8},
\end{equation}
where $\left\langle \phi^8 \right \rangle_{\Sigma_2}$ denotes the $8$ point function.
In the region $t_2 \le l$, that of 8 point function of $\phi$ is given by
\begin{equation}
\left \langle \phi^8 \right \rangle_{\Sigma_2} \sim
\frac{\left(l^2+t_2^2\right) \left(L^2+t_1^2\right)}{262144 \pi ^8 t_2^2 t_1^2 \epsilon ^8}.
\end{equation}

In this limit, the leading term of 8-point function comes from the contributions from only spesific propagators. Their leading term is $\mathcal{O}\left(\epsilon^{-2}\right)$.
In the region $t_2 \leq l$, $\left\langle \phi(r_1, \theta_1)\phi(r_4, \theta_4)\right \rangle_{\Sigma_2 }$, $\left\langle \phi(r_1, \theta_1+2\pi)\phi(r_4, \theta_4+2\pi)\right \rangle_{\Sigma_2 }$, $\left\langle \phi(r_2,\theta_2) \phi(r_3,\theta_3)\right \rangle_{\Sigma_2}$, $\left\langle \phi(r_2,\theta_2) \phi(r_3,\theta_3+2\pi)\right \rangle_{\Sigma_2}$, $\left\langle \phi(r_2,\theta_2+2\pi) \phi(r_3,\theta_3)\right \rangle_{\Sigma_2}$ and $\left\langle \phi(r_2,\theta_2+2\pi) \phi(r_3,\theta_3+2\pi)\right \rangle_{\Sigma_2}$ can contribute to the leading term of 8 point function. In this limit, these propagators are given 
\begin{equation}
\begin{split}
&\left\langle \phi(r_2,\theta_2) \phi(r_3,\theta_3)\right \rangle_{\Sigma_2}=\left\langle \phi(r_2,\theta_2+2\pi) \phi(r_3,\theta_3+2\pi)\right \rangle_{\Sigma_2} \sim \frac{L+t_1}{32 \pi ^2 t_1 \epsilon ^2}+\mathcal{O}(\epsilon^{-1}), \\
&\left\langle \phi(r_2,\theta_2) \phi(r_3,\theta_3+2\pi)\right \rangle_{\Sigma_2}=\left\langle \phi(r_2,\theta_2+2\pi) \phi(r_3,\theta_3)\right\rangle_{\Sigma_2 }= \frac{-L+t_1}{32 \pi ^2 t_1 \epsilon ^2}+\mathcal{O}(\epsilon^{-1}), \\
&\left\langle \phi(r_1, \theta_1)\phi(r_4, \theta_4)\right \rangle_{\Sigma_2 } =\left\langle \phi(r_1, \theta_1+2\pi)\phi(r_4, \theta_4+2\pi)\right \rangle_{\Sigma_2 } =\frac{1}{16 \pi ^2 \epsilon ^2}+\mathcal{O}(\epsilon^0).
\end{split}
\end{equation}

In the region $t_2 \ge l$, the contributions from $\left\langle \phi(r_2,\theta_2) \phi(r_3,\theta_3)\right \rangle_{\Sigma_2}$, $\left\langle \phi(r_2,\theta_2) \phi(r_3,\theta_3+2\pi)\right \rangle_{\Sigma_2}$, $\left\langle \phi(r_2,\theta_2+2\pi) \phi(r_3,\theta_3)\right \rangle_{\Sigma_2}$ and $\left\langle \phi(r_2,\theta_2+2\pi) \phi(r_3,\theta_3+2\pi)\right \rangle_{\Sigma_2}$ do not change. 
On the other hand, the contributions from $\left\langle \phi(r_1,\theta_1) \phi(r_4,\theta_4)\right \rangle_{\Sigma_2}$, $\left\langle \phi(r_1,\theta_1+2\pi) \phi(r_4,\theta_4+2\pi)\right \rangle_{\Sigma_2}$ change as follows,
\begin{equation}
\begin{split}
&\left\langle \phi(r_1,\theta_1) \phi(r_4,\theta_4)\right \rangle_{\Sigma_2}=\left\langle \phi(r_1,\theta_1+2\pi) \phi(r_4,\theta_4+2\pi)\right \rangle_{\Sigma_2}=\frac{l+t_2}{32 \pi ^2 t_2 \epsilon ^2}+\mathcal{O}(\epsilon^{-1}), \\
\end{split}
\end{equation} 
$\left\langle \phi(r_1,\theta_1+2\pi) \phi(r_4,\theta_4)\right \rangle_{\Sigma_2}$ and $\left\langle \phi(r_1,\theta_1+2\pi) \phi(r_4,\theta_4+2\pi)\right \rangle_{\Sigma_2}$ can also contribute to leading term of 8-point function of $\phi$.
These propagator are given 
\begin{equation}
\begin{split}
&\left\langle \phi(r_1,\theta_1) \phi(r_4,\theta_4+2\pi)\right \rangle_{\Sigma_2}=\left\langle \phi(r_1,\theta_1) \phi(r_4,\theta_4+2\pi)\right\rangle_{\Sigma_2} =\frac{-l+t_2}{32 \pi ^2 t_2 \epsilon ^2}+\mathcal{O}(\epsilon^{-1}).
\end{split}
\end{equation} 

The contribution from other propagators is at most $\mathcal{O}(\epsilon^{-1})$. Then they can not contribute to the leading term of 8 ponit function of $\phi$. 

In the $\epsilon \rightarrow 0$ limit, $\Delta S^{(2)}_A$ for this state is given by 
\begin{equation}
\Delta S^{(2)}_A =
\begin{cases}
\log{\left[\frac{2 t_1^2}{L^2+t_1^2}\right]}&  t_2\le l, \\
\log{\left[\frac{4 t_1^2 t_2^2}{\left(l^2+t_2^2\right) \left(L^2+t_1^2\right)}\right]} & t_2 \ge l. 
\end{cases}
\end{equation}

\begin{figure}\label{mL10}	
  \centering
  \includegraphics[width=8cm]{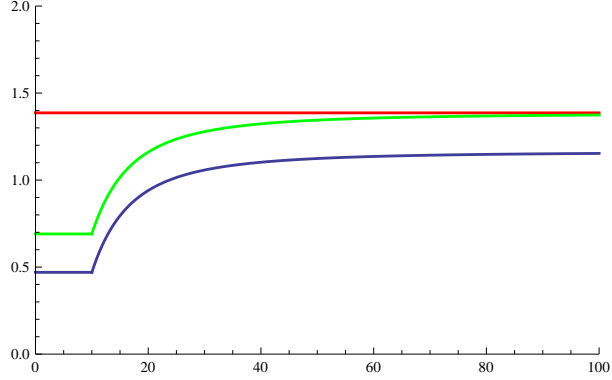}
  \caption{The plots of evolution of $\Delta S^{(2)}_A$ with $t_2$ in the $\epsilon \rightarrow 0$ limit.
The vertical axis corresponds to $\Delta S^{(2)}_A$. The horizontal axis corresponds to $t_2$. Here we chose $l=10$.
The blue curve corresponds to the time evolution of $\Delta S^{(2)}_A$ for $\phi\phi\left|0\right \rangle$, where one of two scalar fields is located at $t=-t_1=-10, x^1=-L=-5$. The green curve corresponds to the time evolution of $\Delta S^{(2)}_A$ for $\phi\phi\left|0\right \rangle$ where one of two scalar fields is located at $t=-t_1=-100, x^1=-L= -5$. 
The red curve corresponds to $2\log{2}$.}
\end{figure}

Its time evolution is plotted in Fig.10. The final value of $\Delta S^{(2)}_A$ is given by 
\begin{equation}
\Delta S^{(2)}_A =\log{2}+\log{\left[\frac{2 t_1^2}{L^2+t_1^2}\right]}
\end{equation}
We are able to control its final value by turning the parameters $t_1$ and $L$.
And the maximum value of it is given by 
\begin{equation}
\Delta S^{(2)\text{max}}_A =2\log{2}.
\end{equation}

\subsubsection{Simultaneous Case}
Let us think about the case where we act two operators $\phi \phi$ simultaneously on the ground state. These operators are located at $x^1=-l$ and $x^1=-L$ respectively. Here, we assume that $L> l$. We investigate the evolution of the second R$\acute{e}$nyi entropy with $t=t_1$ for this state\footnote[6]{If we act one of these operators on the ground state at $t=t_1$and another one on it at $t=t_1+c$~$ (c\ge 0) $, we can investigate the evolution of $\Delta S^{(2)}_A$. Here, we think about the simplest case.}.
We can compute it as we explained above. We just have to change $t_2$ to $t_1$ in (\ref{2ana}). Therefore we skip the detail of the calculation. Of course, we perform the analytic continuation similar to that in (\ref{2ana}) and take the $\epsilon \rightarrow 0$ limit.

The time evolution of second R$\acute{e}$nyi entanglement entropy is given by
\begin{equation}
\Delta S^{(2)}_A = \begin{cases}
0 &0 \le t_1 \le l \\
\log{\left[\frac{2t_1^2}{l^2+t_1^2}\right]} & t_1 \ge l, \\ 
\log{\left[\frac{2t_1^2}{l^2+t_1^2}\right]}+ \log{\left[\frac{2t_1^2}{L^2+t_1^2}\right]} & t_1 \ge L.
\end{cases}
\end{equation}
The evolution of $\Delta S^{(2)}_A$ with $t_1$ is plotted in Fig.11.
If we define the entropy for the state which is excited by $\phi(-t_1,-L)$ and  $\phi(-t_1,-l)$ by $\Delta S^{(2), 1}_A$ and  $\Delta S^{(2), 2}_A$ respectively, then the entropy for the state generated by acting both of these operators on the ground state is given by the sum of these entropies,  
\begin{equation}
\Delta S^{(2)}_A =\Delta S^{(2), 1}_A+\Delta S^{(2), 2}_A
\end{equation}
If we take the $t_1 \rightarrow \infty$ limit, the final value of $\Delta S^{(2)}_A$ is given by 
\begin{equation}
\Delta S^{(2) f}_A=2\log{2}. 
\end{equation} 

\begin{figure}\label{l10L40}	
  \centering
  \includegraphics[width=8cm]{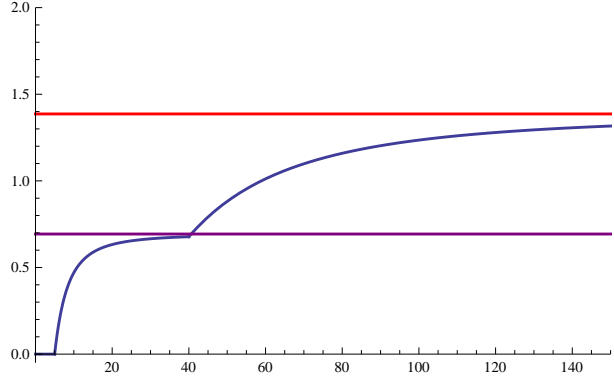}
  \caption{The plots of evolution of $\Delta S^{(2)}_A$ with $t_1$ in the $\epsilon \rightarrow 0$ limit.
The vertical axis corresponds to $\Delta S^{(2)}_A$. The horizontal axis corresponds to $t_1$. Here we chose $l=5$ and $L=40$.
The blue curve is the time evolution of $\Delta S^{(2)}_A$ for $\phi\phi\left|0\right \rangle$ where one of the two scalar fields are spatially located at $x^1=-l=-5$. Another one is located at $x^1=-L=-40$ The purple curve is $\log{2}$. The red curve is $2\log{2}$.
}
\end{figure}

\section{Entangled Pair Interpretation}
In previous section, we investigate the time evolution of $\Delta S^{(n)}_A$.
Here we will interpret the time evolution of $\Delta S^{(n)}_A$ in terms of relativistic propagation of entangled quanta.
Moreover we will acquire $\Delta S^{(n)f}_A$ for the state defined by acting the local operator $:\phi^k:$ on the ground state by using this framework.

We will also explain the new results which we obtain in present paper.
One of them is the large $k$ behavior of $\Delta S^{(n)f}_A$ for the state defined by acting the local operator $:\phi^k:$ on the ground state.
We will find the sum rule for the (R$\acute{e}$nyi) entanglement entropies for the state defined by acting several operators on the ground state.
We assume that these operators are given by $:\phi^{k_i}:$ and located separately.
We will argue $\Delta S^{(n)f}_A$ for the state defined by acting specific operators on the ground state is given by (\ref{gfo}).
We will also argue that the sum rule can be generalized. 
 
\subsection{A Physical Interpretation of $\Delta S^{(n)}_A$ in two dimension}
We investigated the time evolution of $\Delta S^{(n)}_A$ for the locally excited states in 2,4 and 6 dimensional spacetime in previous section.
Here we interpret the time evolutions of them in terms of relativistic propagation of the entangled pair.

Let us consider the result of $\mathcal{N}^{-1}\left(e^{i \alpha \phi}+e^{-i \alpha \phi}\right)\left| 0\right\rangle$ in two dimension in terms of its propagation.
The time evolution of $\Delta S^{(2)}_A$ for this state corresponds to the blue curve in Fig.5.
In this figure, the amount of $\Delta S^{(2)}_A$ can not increase until $t_1=l$.
 It drastically increases at $t_1=l$ and reaches constant value $\log{2}$ at late time.

We can interpret its time evolution in terms of relativistic propagation of entangled quanta as follows.
First, a pair of two quanta appears at the point where that operator is located as in Fig.11.
Here this operator is assumed to be inserted into the complement B of subsystem A.
These two quanta are entangled with each other and we call this pair an entangled pair.
Each of two quanta propagates in the opposite direction at the speed of light.
Until $t_1=l$, two quanta remain to stay in the region B. At that time, the quantum entanglement between them is not able to contribute to $\Delta S^{(2)}_A$.
When one of them reaches at the boundary $\partial A$ of subsystem A, the quantum entanglement between them begins to contribute to the second R$\acute{e}$nyi entanglement entropy between A and B.
Since these quanta propagate at the speed of light, one of them can reach $\partial A$ at $t_1=l$.
We chose a half of total space as the subsystem A.
Therefore, once one of them enters the subsystem A, it can not go out there.
Since each of entangled pair remain to stay in A and B respectively, $\Delta S_A$ can approach a finite constant. In particular, these quanta can propagate in only one direction respective in two dimensional examples. Therefore $\Delta S^{(2)}_A$ suddenly reaches a nontrivial constant at $t_1=l$. In higher dimensional examples($d>1$), those quanta can propagate in generic direction. Then $\Delta S^{(n)}_A$ approaches some constants values mildly as in Fig.7.

\begin{figure}[htbp]
 \begin{minipage}{0.4\hsize}
  \begin{center}
   \includegraphics[width=80mm]{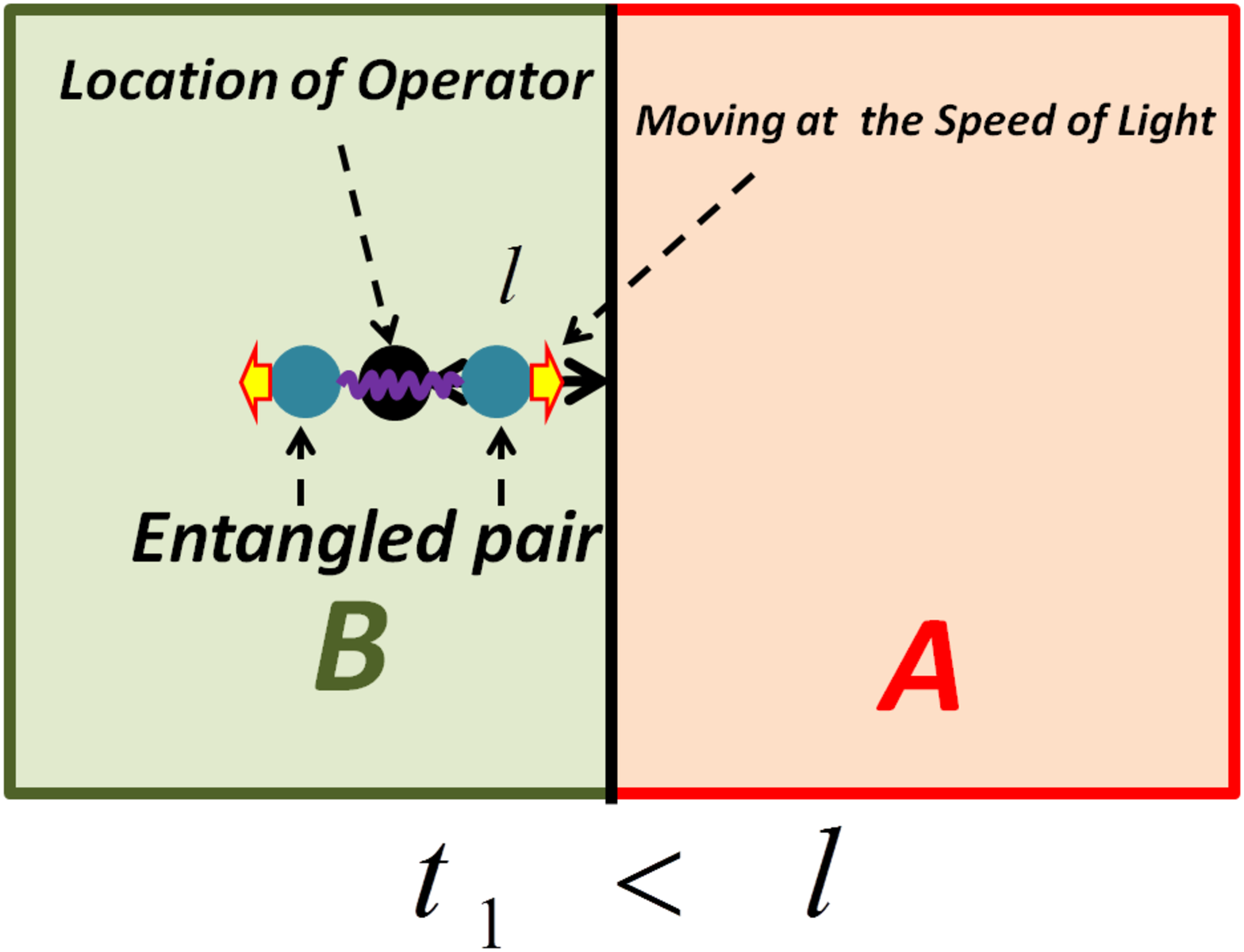}
  \end{center}
 \end{minipage}
  \begin{minipage}{0.8\hsize}
  \begin{center}
   \includegraphics[width=80mm]{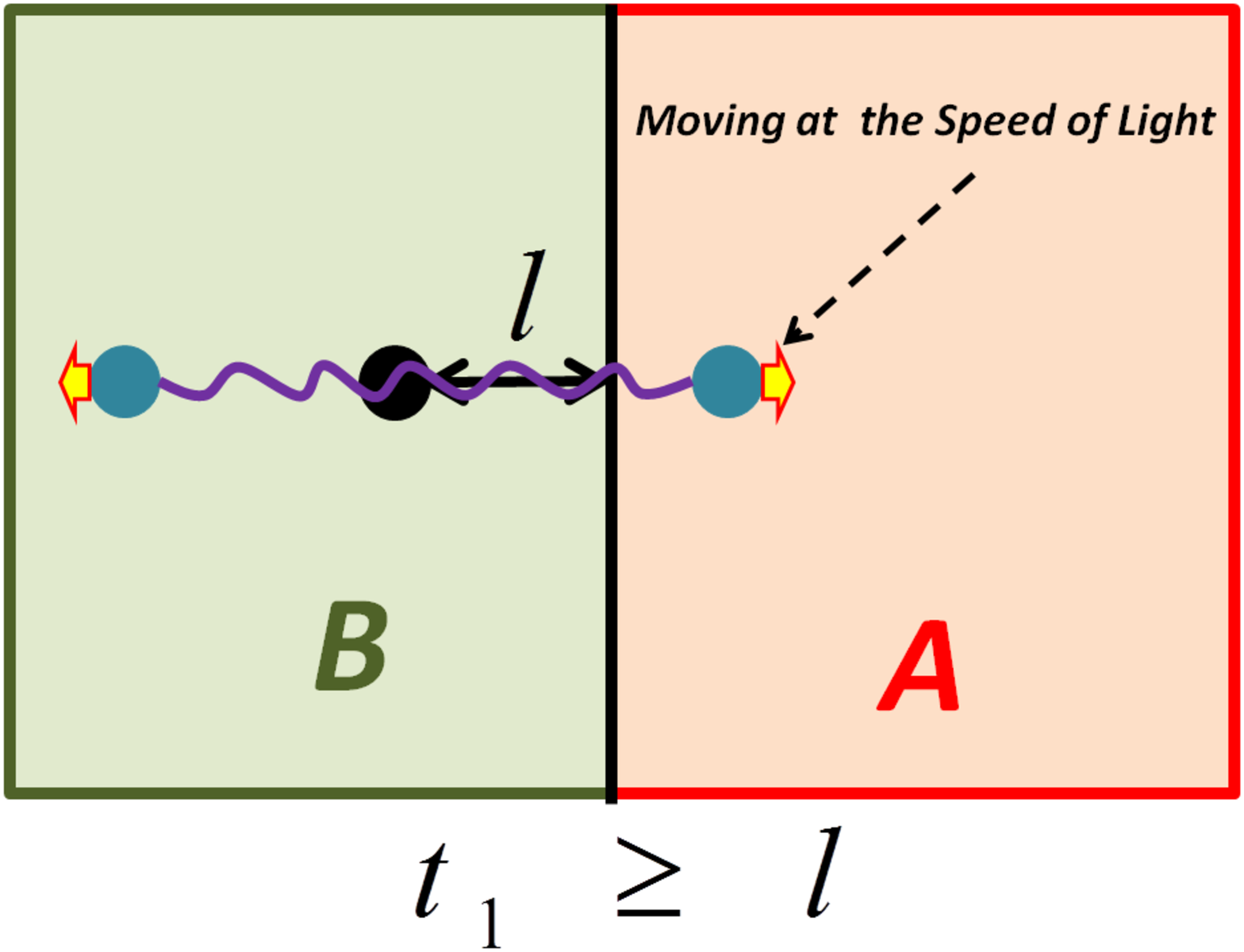}
  \end{center}
 \end{minipage}
 \caption{The schematic explanation for the time evolution of $\Delta S^{(n)}_A$. At $t_1<l$, each of entangled pair is included in region B. At $t_1 \ge l$, two quanta are included in A and B respectively. }
\end{figure}

Moreover we are able to find out why $\Delta S^{(2)f}_A$ is given by $\log{2}$ by using this interpretation.
We decompose $\phi$ into $\phi_L$ and $\phi_R$,
\begin{equation}\label{dec}
\phi =\phi_L+\phi_R,
\end{equation}
where $\phi_L$ and $\phi_R$ respectively correspond to the modes which are moving towards the left direction $(x^1 < 0)$ and the right direction $(x^1>0)$.
Therefore this locally exited state is given by
\begin{equation}
\left|\Psi\right\rangle =\frac{1}{\sqrt{2}}\left| e^{-i\alpha \phi_R}\right \rangle_R \otimes \left|e^{-i\alpha \phi_L}\right \rangle_L +\frac{1}{\sqrt{2}} \left|e^{i\alpha \phi_R}\right \rangle_R \otimes \left|e^{i\alpha \phi_L}\right \rangle_L
\end{equation}
where $\left|\Psi\right \rangle$ is normalized so as to $\left\langle \Psi \big{|} \Psi \right\rangle=1$. This locally excited state is an EPR state (a maximally entangled state). Therefore $\Delta S^{(2)f}_A$ is given by $\log{2}$.

For any $n$, we can acquire $\Delta S^{(n)}_A$ for that state.
Its reduced matrix is given by
\begin{equation}
\rho_{A}=
\frac{1}{2}\left( \begin{array}{cc}
1 & 0 \\
0 & 1 
\end{array} \right),
\end{equation}
and for any $n$, $\rho^n_A$ is given by
\begin{equation}
\rho_{A}^n=
\frac{1}{2^n}\left( \begin{array}{cc}
1 & 0 \\
0 & 1 
\end{array} \right).
\end{equation}

The definition of (R$\acute{e}$nyi) entanglement entropies are given in (\ref{drenyi}).
(R$\acute{e}$nyi) entanglement entropies for this RPR state do not depend $n$. 
Therefore they are given by 
\begin{equation}
\Delta S^{(n)}_A = \log{2}.
\end{equation}

Let us apply this interpretation for other results in two dimensional cases.
In the case of the entropy for the excited state defined by acting the derivatives of $\phi$ mentioned above and $:e^{i \alpha \phi}:\left|0\right\rangle$ on the ground state, these locally excited states can be represented in terms of products states. Therefore $\Delta S^{(n)}_A$ vanishes.
In next subsection, we apply the entangled pair interpretation for the results in higher dimensional cases.

\subsection{A Physical Interpretation in Higher Dimensions $(d>1)$}
Let us apply the entangled pair representation for the results in higher dimensional cases.
And we will obtain $\Delta S^{(n)f}_A$ for the states defined by acting $:\phi^k:$ on the ground state.

We decompose $\phi$ into $\phi_L$ and $\phi_R$ as in $(\ref{dec})$.
Using this decomposition, the excited state defined by acting $:\phi^k:$ on the ground state is given by
\begin{equation}
\left|\Psi \right\rangle =\mathcal{N}^{-1} :\phi^k:\left|0\right\rangle =\mathcal{N}^{-1}\sum^{k}_{m=0} ~_k C_m  \phi_R^m \left|0\right \rangle_R \otimes \phi_L^{k-m} \left|0\right \rangle_L,
\end{equation}
where $\mathcal{N}$ is the normalization factor and$ \left| 0 \right\rangle_R$ and $ \left|0 \right\rangle_L$ are the vacuum for the right moving mode and the left moving mode respectively.
Moreover we define new bases by, 
\begin{equation}
\begin{split}
&\left|j\right\rangle_R =\frac{1}{\sqrt{j !}}(\phi_R)^j\left|0\right\rangle_R, \\
&\left|j\right\rangle_L =\frac{1}{\sqrt{j !}}(\phi_L)^j\left|0\right\rangle_L, \\
\end{split}
\end{equation}
where these states are normalized so as to $\left\langle i\big{|}j \right \rangle_{L, R}= \delta_{i j}$.
And $\left| \Psi \right \rangle$ is also normalized so as to $\left \langle \Psi \big{|} \Psi \right \rangle =1$.
After normalized, this locally excited state is given by 
\begin{equation}
\left|\Psi\right \rangle = \frac{1}{2^{\frac{k}{2}}}\sum^{k}_{m=0}\sqrt{ _k C_m}\left|m\right\rangle_A \otimes \left|k-m\right \rangle_B.
\end{equation}

If we take $t_1 \rightarrow \infty$ limit, we can identify the degrees of freedom of right moving modes and left moving modes with those in subsystem A and B respectively.
Therefore tracing out the degrees of freedom in subsysten B is equivalent to tracing out those of left moving modes.
After tracing out the degrees of freedom in subsystem B, its reduced density matrix is give by
\begin{equation}\label{gmat}
\rho_{A,k} =\frac{1}{2^k}\left( \begin{array}{ccccc}
_k C_0 & 0& 0 &\cdots & 0 \\
0 &_k C_1 & 0 &\cdots & 0 \\
\vdots & \vdots &\ddots & \vdots &\vdots \\
0 & 0 &0 & \cdots & _k C_k \\
\end{array} \right),
\end{equation}
where $_k C_m = \frac{k!}{(k-m)! m!}$ .
It is given by binominal distribution.

By using the matrix in (\ref{gmat}), we can obtain its explicit form for each k as follows,

\begin{equation}\label{mtcs}
\begin{split}
\rho_{A,1}&=
\frac{1}{2}\left( \begin{array}{cc}
1 & 0 \\
0 & 1 
\end{array} \right),
\rho_{A,2}=
\frac{1}{4}\left( \begin{array}{ccc}
1 & 0 & 0\\
0 & 2 & 0\\
0 & 0& 1\\
\end{array} \right), \\
\rho_{A,3}&=
\frac{1}{8}\left( \begin{array}{cccc}
1 & 0 & 0 & 0 \\
0 & 3 & 0 & 0 \\
0 & 0 & 3 & 0 \\
0 & 0 & 0 & 1 \\
\end{array} \right),
\rho_{A,4}=
\frac{1}{16}\left( \begin{array}{ccccc}
1 & 0 & 0 & 0 & 0\\
0 & 4 & 0 & 0 & 0 \\
0 & 0 & 6 & 0 & 0 \\
0 & 0 & 0 & 4 & 0 \\
0 & 0 & 0 & 0 & 1 \\
\end{array} \right), \\
\rho_{A,5}&=
\frac{1}{32}\left( \begin{array}{cccccc}
1 & 0 & 0 & 0 & 0 & 0 \\
0 & 5 & 0 & 0 & 0 & 0\\
0 & 0 & 10 & 0 & 0 & 0\\
0 & 0 & 0 & 10 & 0 & 0\\
0 & 0 & 0 & 0 & 5 & 0\\
0 & 0 & 0 & 0 & 0 & 1\\
\end{array} \right).
\end{split}
\end{equation}

Substituting $\rho_{A, j}$ in (\ref{gmat}) into (\ref{drenyi}), we are able to acquire the $n$-th R$\acute{e}$nyi entanglement entropy for the locally excited state defined by acting $:\phi^k:$ on the ground state,
\begin{equation} \label{gfo}
\Delta S^{(n)f}_{A, k}= \frac{1}{1-n}\log{\left(\frac{1}{2^{n k}}\sum^{k}_m(~_k C_m)^n\right)}.
\end{equation}

It is given by the $n$- th R$\acute{e}$nyi entanglement entropy of binomial distribution.

We can derive the explicit results for each $k, n$ from the formula in (\ref{gfo}). These results do not depend on spacetime dimension as long as its dimension is higher than $2$.
As you see later, they are consistent with results which we will obtain by the replica method.
These results are summarized in Table.1. In any dimensions, the final values of the (R$\acute{e}$nyi) entanglement entropies for the state which is excited by $:\phi^k:$ are given by (\ref{gfo}). 

In $n=2$ case, the formula in (\ref{gfo}) is simplified,
\begin{equation}
\Delta S^{(2) f}_{A, k} =\sum^{k}_{m=1}\log{\frac{2m}{2m-1}}
\end{equation}
$\Delta S^{(n)}_{A,1}$ does not depend on the number of sheets. It is equivalent to the (R$\acute{e}$nyi) entanglement entropy for an EPR states (maximally excited state).
Using the formula in (\ref{gfo}), we can compute $\Delta S^{(n)f}_{A, k}$ for the state defined by acting $:\phi^k:$ on the ground state.
As we will see later, the formula in (\ref{gfo}) gives a characterization of operators. 

We define the (R$\acute{e}$nyi) entanglement entropies of operators by the final value of (R$\acute{e}$nyi) entanglement entropies for the state defined by acting them on the ground state. It characterizes the quantum entanglement of operators and it is independent of the conformal dimension of operators\footnote[7]{In two dimensional cases, We have seen that $\Delta S^{(n) f}_A $ does not depend on conformal dimension of operators. The conformal dimensions of $\mathcal{O}_1=:e^{i \alpha \phi}:$ is the same as the conformal dimension of $\mathcal{O}_2= :e^{i \alpha \phi}:+:e^{-i \alpha \phi}:$. However $\Delta S^{(n)f}_A$ for $\mathcal{O}_1$ is not the same as that for $\mathcal{O}_2$.}.
 The quantum entanglement of operators is related to the number of entangled pair which those operators are able to create at least in free massless scalar field theory. 

\begin{table}[ttt]
\caption{The value of $\Delta S^{(l) f}_{A,m}$ for free massless scalar field theories in dimensions higher than two. Here $n$ is the number of sheet. $m$ is the number of $\phi$}
  \begin{tabular}{|c|c|c|c|c|c|c|} \hline
 Entropy   & $l$ & $m=1$ & $m=2$ & $m=3$ & $\cdots$ & $m=j$\\ \hline \hline
      & $2$ & $\log{2}$ & $\log{\frac{8}{3}}$ & $\log{\frac{16}{5}}$  & $\cdots$ & $-\log{\left(\frac{1}{2^{2j}}\sum^j_{i=0}\left(_j C_i\right)^2\right)}$ \\ \cline{2-7}
   $ \Delta S_{A, m}^{(l)f} $ & $3$ & $\log{2}$ & $\frac{1}{2}\log{\frac{32}{5}}$ & $\frac{1}{2}\log{\frac{64}{7}}$  & $\cdots$ & $ \frac{-1}{2}\log{\left(\frac{1}{2^{3j}}\sum^j_{i=0}\left(_jC_i\right)^3\right)}$\\ \cline{2-7}
      & $\vdots$ & $\vdots$ & $\vdots$ & $\vdots$ & $\vdots$& $\vdots$\\ \cline{2-7}
& $n$ & $\log{2}$ & $\frac{1}{n-1}\log{\frac{2^{2n-1}}{2^{n-1}+1}}$ & $\frac{1}{n-1}\log{\frac{2^{3n-1}}{3^n+1}}$ &  $\cdots$ & $\frac{1}{1-n}\log{\left(\frac{1}{2^{n  j}}\sum_{i=0}^{j}\left(_j C_i\right)^{n}\right)}$ \\ \hline
    $ \Delta S_{A,j}$ & $1$ & $\log{2}$ & $\frac{3}{2}\log{2}$ &$\frac{3}{4}\log{\frac{16}{3}}$ & $\cdots$ & $\displaystyle j\cdot \log{2}-\frac{1}{2^j}\sum_{i=0}^{j}~_jC_i\log{ ~_jC_i}$ \\ \hline
  \end{tabular}
\end{table}

\subsubsection{Large $k$ Behavior}
Here we explain one of the new results.
We would like to comment on the large $k$ behavior of $\Delta S^{(n)}_A$.
The (R$\acute{e}$nyi) entanglement entropies of $:\phi^k:$ are given by those of binomial distribution. Therefore in large $k$ limit we can approximate its reduced density matrix by normal distribution as follows
\begin{equation}
\rho_A =\frac{_k C_x}{2^k} \sim \sqrt{\frac{a}{k\pi}}e^{-\frac{a \left(x-\frac{k}{2} \right)^2}{k}},
\end{equation}
where $a$ is a certain positive constant.
By using this approximation, we would like to evaluate $\Delta S^{(n)}_{A, j}$ in Table.1. 
 $\Delta S^{(n)}_{A, j}$ are given by 
\begin{equation}
\Delta S^{(n)}_A \sim \frac{1}{2}\log{k}.
\end{equation} 
They do not depend on the number of sheets.
\subsubsection{Generalization of $\Delta S^{(n) f}_{A, j}$}

We explained the (R$\acute{e}$nyi) entanglement entropies of $:\phi^k:$ up to here. Let us consider the (R$\acute{e}$nyi) entanglement entropy for the states defined by acting specific operators on the ground state here. We limit operators to particular operators such as $:(\partial^l\phi)^k:$.
They are composed of single-species operators such as $\phi$, $\partial \phi$, $\partial \partial \phi$ and so on.
And we call these operators S-operators.
In the argument above, we assumed that operators can be decomposed to left moving modes and right moving modes at late time. In this assumption, we did not use the detailed feature of local operator $\phi$. Therefore, we argue that we are able to decompose these operators $(\phi$, $\partial \phi$, $\partial \partial \phi, \cdots)$ into the right moving mode and left moving mode when states are excited by S-operators. Therefore in free massless scalar field theory in any dimension, the (R$\acute{e}$nyi) entanglement entropies of S-operators are given by the formula in (\ref{gfo}).

In summary, we argue that the (R$\acute{e}$nyi) entanglement entropies of S-operators should be given by (\ref{gfo}).
This implies that quantum entanglement of operators is independent of the conformal dimensions of operators. The (R$\acute{e}$nyi) entanglement entropies of operators  characterize  operators in the viewpoint of quantum entanglement\footnote[8]{The authors in \cite{ttq} found the entanglement of operator are related to quantum dimension in rational CFTs on $2$ dimensional space}.
We have explicitly computed the quantum entanglement of $(\partial \phi)^j$, for $n=2, 3$ in $4 dimension$. We found that they are consistent with the results we obtained under this assumption.
Moreover, $\Delta S^{(n) f}_A$ which we derive from the entangled pair interpretation  agrees with the result which we will obtain by the replica method.
\subsection{Sum Rule of (R$\acute{e}$nyi) entanglement entropies of Operators}
We will interpret the final value of (R$\acute{e}$nyi) entanglement entropy for states  by acting various operators on the ground state in terms of entangled pair.
First, we think about $\Delta S^{(n)}_A$ for $\mathcal{N}^{-1}\mathcal{T}:\phi_1^{m_1}: :\phi_2^{m_2} :\left|0\right\rangle$. These operators are located separately.
We assume $:\phi_1^{m_1}:$, $:\phi_2^{m_2}:$ can be independently decomposed to left moving modes and right moving modes as follow,
\begin{equation}
\begin{split}
&\phi_1 =\phi_{L,1}+ \phi_{R, 1},\\
&\phi_2 =\phi_{L, 2}+ \phi_{R, 2}, \\
\end{split}
\end{equation}
where $\phi_{L, 1}$ and $\phi _{R,1}$ are independent of other operators $\phi_{L, 2}, \phi_{R, 2}$.
Then the annihilation and creation operators of $\phi_{L, R, 1}$ can commute with those of $\phi_{L, R, 2}$\footnote[9]{If these two operators are located on the light cone. It is not clear that they are independent of each other. However in the replica method, it does not seem $\Delta S^{(n)}_A$ shows nontrivial behavior if two operators are located on the light cone.}.

Under this assumption, the normalized excited state is given by
\begin{equation}
\begin{split}
&\left| \Psi \right \rangle =\mathcal{N}^{-1} \mathcal{T} :\phi^{m_1}_1: :\phi^{m_2}_2:\left|0\right\rangle \\
&~~~~=\frac{1}{2^{\frac{m_1}{2}}2^{\frac{m_2}{2}}}\sum^{m_1}_{k_1=0}\sum^{m_2}_{k_1=0}\sqrt{~_{m_1}C_{k_1}~_{m_2}C_{k_2}}\left| m_{1}-k_{1}, m_{2}-k_2 \right\rangle_L \otimes \left|k_1, k_2 \right \rangle_R,
\end{split}
\end{equation}
where $\mathcal{T}$ is the time ordering operator.

In any dimensions, the final value of the $n$-th R$\acute{e}$nyi entanglement entropy for this state is given by
\begin{equation} \label{2sum}
\Delta S^{(n) f}_A = \Delta S^{(n) f}_{A, m_1}+\Delta S^{(n) f}_{A, m_2}
\end{equation}
where $\Delta S^{(n) f}_{A,m_1}$ and $\Delta S^{(n) f}_{A,m_2}$ are given by (\ref{gfo}). They are the (R$\acute{e}$nyi) entanglement entropy of $:\phi^{m_1}:$ and $:\phi^{m_2}:$ respectively.
$\Delta S^{(n) f}_A$ is given by the sum of  $\Delta S^{(n) f}_{A,m_1}$ and $\Delta S^{(n) f}_{A,m_2}$.
When $m_1$, $m_2$ are $1$ respectively in $n=2$ case, $\Delta S^{(2) f}_{A}$ is consistent with the result which we obtained in previous section. 

The sum rule in (\ref{2sum}) can be generalized. 
We prepare the locally excited state defined by acting several operators on the ground state.
 These operators $\mathcal{O}^i$ $(i =1 \sim q)$ are given by $:\phi^{k_i}:$. 
We assume they are located separately. In any dimensions $(d>1)$, $\Delta S^{(n) f}_A$ at late time for the state generated by acting these operators on the ground state are given by 
\begin{equation}\label{gsum}
\Delta S^{(n) f }_A =\sum_{i}^q \Delta S^{(n) f i}_{A, k_i},
\end{equation}
where $\Delta S^{(n) f i}_A$ is the (R$\acute{e}$nyi) entanglement entropy of the local operator $\mathcal{O}^i$.
As you see later, the result in (\ref{gsum}) agrees with the result which we obtain by replica method.

\subsubsection{Generalized Sum Rule}
We would like to generalize the sum rule of (R$\acute{e}$nyi) entanglement entropies of operators to those in more general case where operators are not limited to $:\phi^{k_i}:$.
We did not use the distinctive feature of $\phi$ when we found the sum rule of $\Delta S^{(n) f}_A$.
We assumed that $\phi$ can be decomposed to left moving mode and right moving mode.
If the location of $\phi_1$ is separated from the location of $\phi_2$, $\phi_1$ is independent of $\phi_2$.
It is expected that we can apply this assumption for the (R$\acute{e}$nyi) entanglement entropy of operators other than $:\phi^k:$. 
Therefore we argue that if the state is defined by acting various general operators $\mathcal{O}^{i}$ on the ground state, $\Delta S^{(n) f}_A$ for that state should obey the sum rule which is given by (\ref{gsum}). Here, we assume that $\mathcal{O}^{i}$ are located separately. In other words, $\Delta S^{(n) f}_A$ for that state is given by the sum of the (R$\acute{e}$nyi) entanglement entropies of those operators in any dimensions.

In summary, the excited state is given by 
\begin{equation}
\left| \Psi \right \rangle = \prod_{i=1}^q \mathcal{O}^i(t_i, x_i)\left|0\right \rangle,
\end{equation}
where the locations of operators are separated from the locations of operators other than itself. We assume that $\mathcal{O}^i$ are general operators.

In any dimensions $(d>1)$, the (R$\acute{e}$nyi) entanglement entropy for this excited state is given by 
\begin{equation} 
\Delta S^{(n) f }_A =\sum_{i=1}^q \Delta S^{(n) f i}_A,
\end{equation}
where $\Delta S^{(n) f i}$ is the (R$\acute{e}$nyi) entanglement entropy of $\mathcal{O}^i(t_i, x_i)$.

\section{General Argument Using Propagators}
In this section we would like to explain the new results which we obtain in present paper. Firstly we find the explanation which shows that the (R$\acute{e}$nyi) entanglement entropies of$:\phi^k:$ are given by those of binomial distribution by the replica method. And we will also obtain the sum rule which the (R$\acute{e}$nyi) entanglement entropies for the state defined by acting various operators on the ground state. We assume that these operators are located separately. Moreover we argue that the (R$\acute{e}$nyi) entanglement entropies of the specific operators $\left( \partial^m \phi\right)^n$ are given by the (R$\acute{e}$nyi) entanglement entropies of binomial distribution. We will also argue that $\Delta S^{(n)}_A$ for the state defined by acting various operators on the ground state obey the same sum rule. These results which we obtain by replica trick will agree with the results which we obtained in terms of entangled pair. Finally we will comment on the effects of a conical singularity. On $\Sigma_n$ there is a conical singularity on the boundary of the subsystem A. At first, it seems that this can affect the R$\acute{e}$nyi entanglement entropy for locally excited states. However it can not affect the entropy for those states as we will explain later. 

\subsection{The Properties of Propagators} 
In the free scalar field theories, we have to calculate propagators to compute the correlation functions of local operators.
By taking the $\epsilon \rightarrow 0$ limit, the calculations of $\Delta S^{(n)}_A$ are simplified due to distinctive features of propagators.
We assume the operators located at $(r_1, \theta_1)$ and $(r_2, \theta_2)$ on $\Sigma_1$ as in Fig.1.

\begin{figure}[htbp]
 \begin{minipage}{0.4\hsize}
  \begin{center}
   \includegraphics[width=80mm]{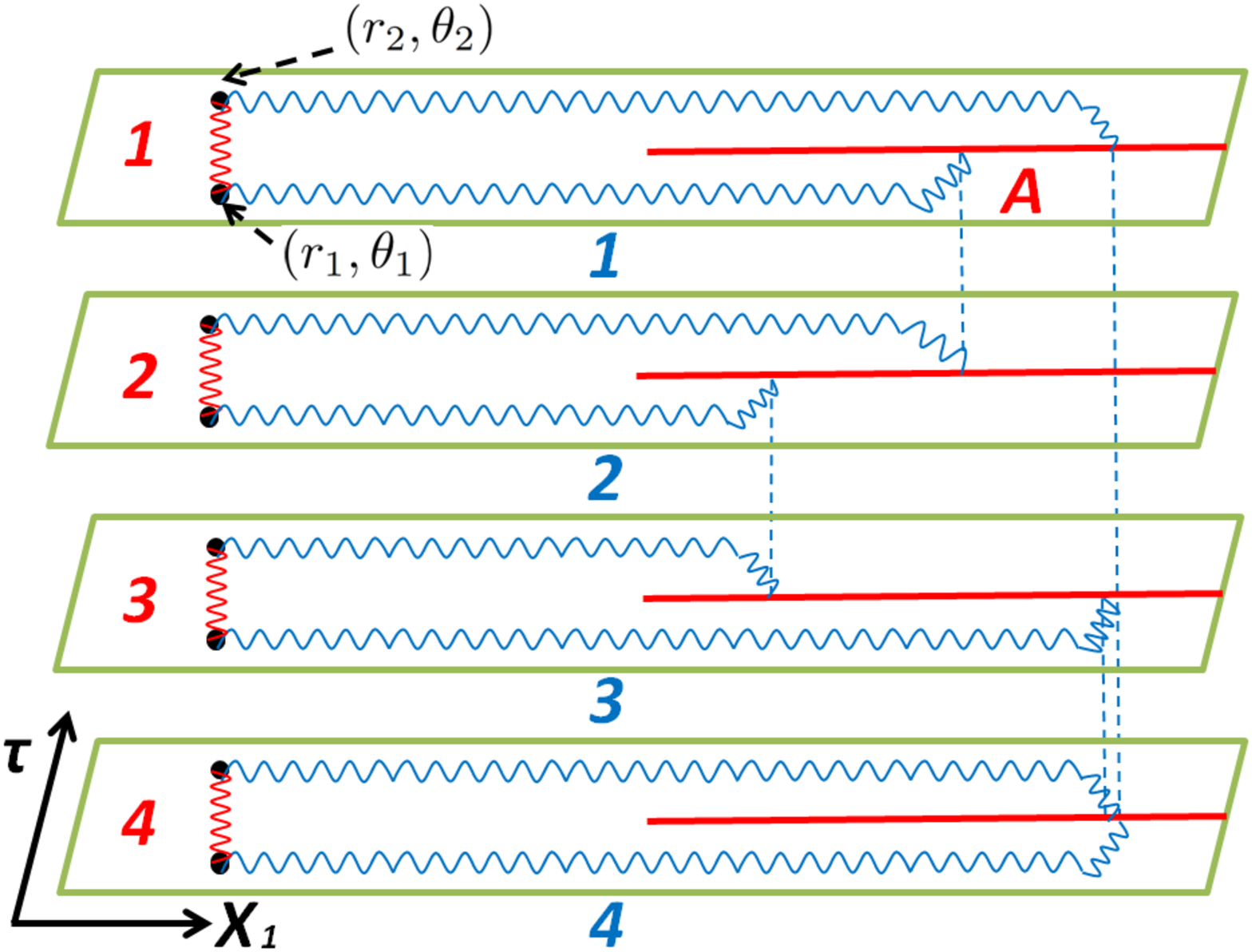}
  \end{center}
 \end{minipage}
  \begin{minipage}{0.8\hsize}
  \begin{center}
   \includegraphics[width=80mm]{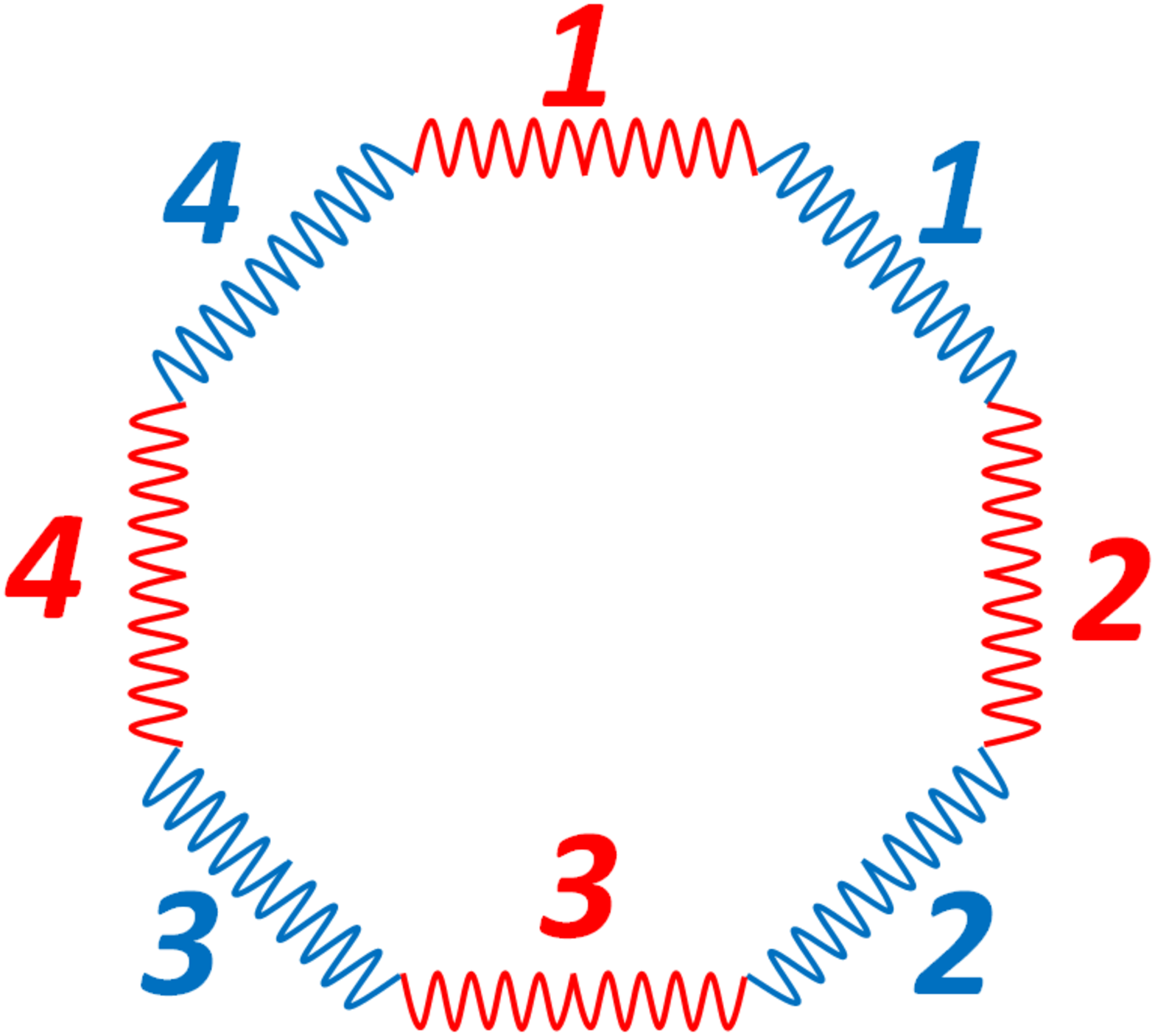}
  \end{center}
 \end{minipage}
 \caption{The circle of dominant propagators on $4$ sheeted geometry. There are $4$ blue wavy lines and red wavy lines respectively. The dots on $\Sigma_4$ are the points where operators are located}
\end{figure}

 Only specific  propagators on $\Sigma_n$ can dominantly contribute to $2n$ point correlation function of local operators. We define these dominant propagators on $\Sigma_n$ by $D^{(n)}$.  As in Fig.13, the propagators which are aligned on a circle are dominant propagators. In this figure, it shows the configuration of propagators on $\Sigma_4$. This circle is made of $2n$ propagators. And these $2n$ two point functions necessarily include the propagators between two points on same sheet. These propagators respectively correspond to the $4$ red wavy lines in Fig.13.
And the other dominant propagators are those between $(r_1, \theta_{1, k})$ and $(r_2, \theta_{2, k+1})$ ~$( k=1, \cdots, n, \theta_{2,n+1}=\theta_{2}).$
They respectively correspond to the $4$ blue wavy lines in Fig.13. The other propagators on $\Sigma_n$ can not contribute to $2n$ point functions. Here we assume that operator is located at $\left(t=-t_1, x^1= -l\right)~(t_1>l)$. In other words, we explain the features of propagators in the region where $\Delta S^{(n)}_A$ increase nontrivially.

 Even if we take the $\epsilon \rightarrow 0$ and take $t\rightarrow \infty$, then these dominant propagators can contribute to $\Delta S^{(n)}_A$. The ratios of  $D^{(n)}$ to $D^{(1)}$ does not depend on the number of sheets and dimension of spacetime. This ratio is given by 
\begin{equation}\label{apro}
\begin{split}
\frac{D^{(n)}}{D^{(1)}}&=\frac{ \text{the number of sheets} }{\text{the number of propagators on the circle }
   } =\frac{n}{2 n}=\frac{1}{2}.
\end{split}
\end{equation}

By using this property of dominant propagators, we can acquire the final value $\Delta S^{(n) f}_A$  for $:\phi^k:$ in any dimensions.  
There are two contributions to the $2n$ point correlation function of $:\phi^k:$ on $\Sigma_n$.
 One contribution corresponds to the product of propagators between two points on same sheet.
This corresponds to the red wavy lines in the right figure of Fig.13.
Another one corresponds to the product of propagators between $(r_1, \theta_{1, k})$ and $(r_2, \theta_{2, k+1})$.
This corresponds to the blue wavy lines in the right figure of Fig.13.
Then this $2n$ point function of $:\phi^k:$ on $\Sigma_n$ is given by
\begin{equation}
\begin{split}
&\left\langle :\phi^k (r_1,\theta_1): :\phi^k(r_2,\theta_2):\cdots :\phi^k(r_1,\theta_1+2\pi(n-1)): :\phi^k(r_2,\theta_2+2\pi(n-1)): \right\rangle  \\
&= (k !)^n \left(\frac{D^{(1)}}{2}\right)^{k n}+(k ! _k C_1)^n \left(\frac{D^{(1)}}{2}\right)^{k n}+ \cdots +(k !)^n \left(\frac{D^{(1)}}{2}\right)^{k n} \\
&=\sum^{k}_{l}\left( k ! _k C_l \right)^n \left(\frac{D^{(1)}}{2}\right)^{k n}
\end{split}
\end{equation}

The denominator $\left\langle :\phi^k: :\phi^k:\right \rangle^n$ is given by
\begin{equation}
\left\langle :\phi^k:(r_1,\theta_1) :\phi^k: (r_2,\theta_2)\right \rangle ^n= \left( k!\left(D^{(1)}\right)^k\right)^n. 
\end{equation}

They are proportional to $\left( D^{(1)}\right)^{n k}$ respectively. Then $D^{(1)}$ is cancelled out when we compute $\Delta S^{(n)}_A$. Therefore $\Delta S^{(n) f}_A$ does not depend on the detail of $2$ point function of $\phi$.
In any dimensions, $\Delta S^{(n)f}_A$ is given by 
\begin{equation} \label{rgfo}
\Delta S^{(n)f}_A =\frac{1}{1-n}\log{\left[\frac{1}{2^{n k}}\sum^{k}_{l=1}\left( _k C_l\right)^n\right]}.
\end{equation}
In previous paper \cite{pne}, we obtained the (R$\acute{e}$nyi) entanglement entropies of $:\phi^k:$ in terms of entangled pair. However we did not obtain them for any k and n by replica method. In present paper, we obtain it for any k and n by using the distinctive property of propagators. Moreover the result in (\ref{rgfo}) agrees with that which we obtained in terms of the entangled pairs.  The relation between $D^{(n)}$ and $D^{(1)}$ does not depend on the details of the propagators. Therefore, it is expected that that relation can be generalized to that for propagators of general fields in free massless scalar field theory.

\subsubsection{Time evolution of R$\acute{e}$nyi entropy for excited states}
We would like to comment on the time evolution for $\Delta S^{(n)}_A$ for the states defined by acting $:\phi^k:$ on the ground state. Here we assume that operator is located at $\left(t=-t_1, x^1,=-l\right)~(t_1>l)$. In $4$ dimension, we are able to find the time dependence of the dominant propagators. Its time dependence does not depend on the number of sheets in the $\epsilon \rightarrow 0$ limit. 
In the $\epsilon \rightarrow 0$ limit,  the dominant propagators between two points on the same sheet is given by
\begin{equation}
D^{(n)}_s=\frac{l+t_1}{32\pi^2t_1\epsilon^2 }+\mathcal{O}(\epsilon^0).
\end{equation}
In this limit, the dominant propagators between $(r_1, \theta_{1,k})$ and $(r_2. \theta_{2, k+1})$ is given by
\begin{equation}
D^{(n)}_a=\frac{-l+t_1}{32\pi^2t_1\epsilon^2 }+\mathcal{O}(\epsilon^0).
\end{equation}

We can investigate the time evolution of the (R$\acute{e}$nyi) entanglement entropies for the states excited by $:\phi^k:$.
The growth of the entropy for this state is given by 
\begin{equation}
\Delta S^{(n)}_A = \frac{1}{1-n}\log{\left[\left(16\pi^2\epsilon^2\right)^{k n}\sum_{i=0}^k(_kC_i)^n(D^{(n) }_a)^{n i}(D^{(n) }_s)^{n (k-i)}\right]}~~~~ t \ge l,
\end{equation}
where $\Delta S^{(n)}_A$ does not depend on $\epsilon$ in $\epsilon \rightarrow 0$ imit.

\subsection{Sum Rule of (R$\acute{e}$nyi) Entanglement Entropies of Local Operators}

We point out a new relation called sum rule by the replica method. We define the particular operators $\mathcal{O}^k_i$ by $:\phi^{k_i}:$. Here $i$ means the position of composite operators and $k$ means the powers of $\phi$. Operators $\mathcal{O}^{k_1}_{i_1},\mathcal{O}^{k_2}_{i_2},\cdots, \mathcal{O}^{k_l}_{i_l}$ act on the ground state. These operators are located separately. 
We can find out the final value of $\Delta S^{(n)}_A$ in the $\epsilon \rightarrow 0$ limit by the replica method.

In present case there are also the correlations between operators which are located separately. At first, it seems that they can contribute to the (R$\acute{e}$nyi) entanglement entropies for the locally excited states. However, only dominant propagators can contribute to $\Delta S^{(n)}_A$ if we take the  $\epsilon \rightarrow 0$ limit. In this limit, the leading term of dominant propagators are given by 

\begin{equation}
D^{(n)}\sim \mathcal{O}\left(\left(\epsilon^{-2}\right)^{\frac{d-1}{2}}\right),
\end{equation} 
where $d>2$.

On the other hand, the leading term of the propagators between operators which are located separately is much smaller than that of dominant propagators. Therefore these correlation functions can not contribute to $\Delta S^{(n)}_A$. We are able to explain why only specific propagator can contribute to $\Delta S^{(n)}_A$ as follows.
 Here we define the replica fields of $\phi (r_1, \theta_1)$ and its copy $\phi (r_2, \theta_2)$ on each sheet $i$ by  $\phi(r_1, \theta_{1, i})$ and  $\phi(r_2,\theta_{2, i})$ respectively $(i =1 \cdots n)$. On $\Sigma_1$, $\epsilon$ is a regularization parameter for the correlation function of $\phi(r_1, \theta_1)$ and its copy $\phi(r_2, \theta_2)$ as in Fig.14. The distance between $\phi (r_1, \theta_{1, i})$ and $\phi(r_2, \theta_{2, j})$  greatly depends on the parameter $\epsilon$.
On the other hand, the distance between $\phi (r_1, \theta_{1, i})$ and $\phi(r_3, \theta_{3, j})$  does not depend on $\epsilon$ compared to the distance between $\phi (r_1, \theta_{1, i})$ and $\phi(r_2, \theta_{2, j})$.
Here $\phi(r_3, \theta_{3})$ is put on the different point from the location of $\phi(r_1, \theta_1)$ on $\Sigma_1$. 
The distance between $\phi (r_1, \theta_{1, i})$ and $\phi(r_2, \theta_{2, j})$ shrinks compared to the distance between $\phi (r_1, \theta_{1, i})$ and $\phi(r_3, \theta_{3, j})$. Therefore only dominant propagators can be greatly enhanced compared to the other propagators when we take $\epsilon \rightarrow 0$ limit.
In any dimensions, $\Delta S^{(n)}_A$ is given by the sum of the (R$\acute{e}$nyi) entanglement entropy $\Delta S^{(n), m}_A $ of each operators, 
\begin{equation}
\Delta S^{(n)}_A = \Delta S^{(n), 1}_A  +\Delta S^{(n), 2}_A + \cdots +\Delta S^{(n), q}_A =\sum^{q}_{i=1}\Delta S^{(n), i}_A,
\end{equation}
and $\Delta S^{(n) f}_A $ also obey the similar relation, 
\begin{equation}
\Delta S^{(n) f}_A = \Delta S^{(n) f, 1}_A  +\Delta S^{(n)f, 2}_A + \cdots +\Delta S^{(n) f, q}_A =\sum^{q}_{i=1}\Delta S^{(n) f, i}_A.
\end{equation}

\begin{figure}\label{reg}	
  \centering
  \includegraphics[width=80mm]{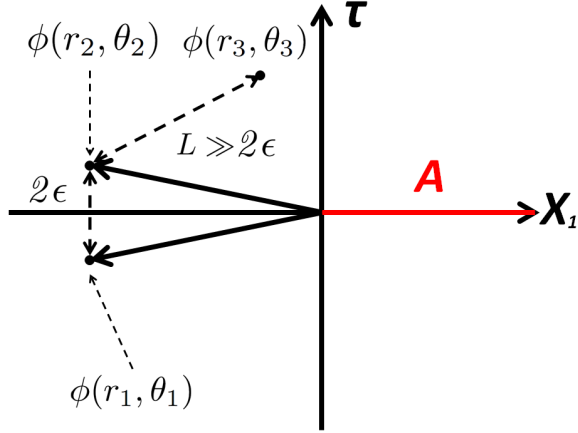}
  \caption{ The locations of operators on $\Sigma_1$. The distance $L$ between $\phi (r_1, \theta_{1})$ and $\phi(r_3, \theta_{3})$ is much larger than the distance $2\epsilon$ between $\phi (r_1, \theta_{1})$ and $\phi(r_2, \theta_{2})$.}
\end{figure}

\subsection{Generalization of Sum Rule}

In previous subsections, we obtained the (R$\acute{e}$nyi) entanglement entropies of $:\phi^k:$ and the sum rule which they obey in the replica method. Only dominant propagators can contribute to them and the computation of $\Delta S^{(n)}_A$ is very simplified in the $\epsilon \rightarrow 0$ limit. In that argument about propagators, we do not use the property of the propagators of $\phi$.

Therefore we argue that the propagators of derivatives of scalar such as $\partial ^k \phi$ which are aligned on the circle dominantly contribute to the $2n$ point correlation functions of them. We define the propagators of these fields on $\Sigma_n$ by $G^{(n)}$. These propagators are given by
\begin{equation}
G^{(n)}= \frac{\text{the number of sheets}}{\text{the number of propagators on cirle}}G^{(1)}=\frac{n}{2n}G^{(1)} =\frac{1}{2} G^{(1)}.
\end{equation}

If the excited state is defined by acting operators  which composed of only single-species operator such as $:(\partial^m \phi)^k:$ on the ground state, then the final value of that entropy for these excited states is given by (\ref{rgfo}) in any dimensions.  
For example, if the local operator is $:\left(\partial_r \phi \right)^k:$, then the final value of  $n$-th R$\acute{e}$nyi entanglement entropy is given by

\begin{equation}\label{rphi}
\Delta S^{(n)f}_A =\frac{1}{1-n}\log{\left[\frac{1}{2^{n k}}\sum^{k}_{l=0}\left( _k C_l\right)^n\right]}.
\end{equation}

We calculate $\Delta S^{(2,3)}_A$ for the state defined by acting local operators $:\left(\partial_r \phi \right)^k:$ on the ground state directly and we check they agrees with the result in (\ref{rphi}).
Furthermore the results which we obtained by the replica method are consistent with that in (\ref{rphi}).

In turn we prepare the excited state defined by acting various operators on the ground state such as 
\begin{equation}\label{mst}
\left|\psi\right\rangle =\mathcal{N}^{-1}\mathcal{T}\prod_{i}^k \mathcal{O}^i(t^i,x^{1,i})\left|0\right\rangle
\end{equation}
where $\mathcal{O}^i$ are general local operators and they are located separately. The index $i$ denotes the location of operators.

In this case, it seems that the correlation between $\mathcal{O}^a(t^a, x^{1, a})$ and $\mathcal{O}^b(t^b, x^{1, b}) ~(a\neq b)$ can contribute to $\Delta S^{(n)}_A$. However the magnitudes of these propagators are also much smaller than those of dominant propagators when we take the $\epsilon \rightarrow 0$ limit. 
Therefore we argue that in any dimensions, (R$\acute{e}$nyi) entanglement entropy for the state in (\ref{mst}) should obey 
\begin{equation}\label{msrule}
\Delta S^{(n) }_A =\sum^{q}_{m}\Delta S^{(n),m}_A.
\end{equation}
where $\Delta S^{(n),k}_{A}$ is the entropy for the state excited by $\mathcal{O}^k$.
And the final value of these entropy also obey the similar rule to that in (\ref{msrule}).

\subsection{Conical Singularity}
We constructed the $n$-sheeted geometry by gluing region A on a sheet to the subsystem A on next sheet as in Fig.2.
There is a conical singularity on the boundary $\partial A$ of the region A on the n-sheeted geometry. 
In our case, it is located at $(\tau = 0, x^1 = 0)$. 
This conical singularity deforms the action of the free massless scalar field in more than $2$ dimension. It adds the conformal mass $c_1 \int dx^{d+1} \mathcal{R}\phi^2$ to that action,
\begin{equation}
S_{Def}=\int dx^{d+1} \frac{1}{2}\partial_{\mu}\phi \partial^{\mu} \phi + c_1 \mathcal{R} \phi^2 ,
\end{equation}
where $c_1$ is a some constant. 
This conformal mass can make an effect on (R$\acute{e}$nyi) entanglement entropies if the state is excited\footnote[10]{The authors in \cite{LMal} pointed out this effect.}.
Therefore we have to take its effect into account. We treat this conformal mass as operator.
If we take the effect of the conical singularity in account, the $2n$ point correlation function of operators on $\Sigma_n$ is given by
\begin{equation}\label{coni}
\begin{split}
\left \langle \mathcal{O}^{2n} \right \rangle_{\Sigma_n^{con}}&=\left \langle \mathcal{O}^{2n}: e^{c_1 \int dx^{d+1}\mathcal{R}\phi^2}: \right \rangle_{\Sigma_n} \\
&=\sum_{k=0}^{\infty}\left \langle \mathcal{O}^{2n} :\frac{\left( c_1 \int dx^{d+1}\mathcal{R}\phi^2\right)^k} {k!}:\right \rangle_{\Sigma_n} \\
&=\left \langle \mathcal{O}^{2n} \right \rangle_{\Sigma_n}+\sum_{k=1}^{\infty}\left \langle \mathcal{O}^{2n} :\frac{\left( c_1 \int dx^{d+1}\mathcal{R}\phi^2\right)^k}{k!}: \right \rangle_{\Sigma_n}.
\end{split}
\end{equation}

The second term of the last line in (\ref{coni}) comes from the conical singularity. In this term, some of $2n$ operators has to contract with the operators at $(\tau = 0, x^1=0)$.
The propagators between the operator at origin and operators at other point weakly depend on $\epsilon$ compared to dominant propagators. 
Therefore when we take $\epsilon \rightarrow 0$ limit, the second term in (\ref{coni}) can not contribute to leading term of $2n$ point correlation function of $\mathcal{O}$. We can ignore the effect of the conical singularity on the boundary of the region A.

By using the simplest example, we explain why the conical singularity can not contribute to the $2n$ point function of $\mathcal{O}$.
We compute the second R$\acute{e}$nyi entanglement entropy for $\mathcal{N}^{-1} \phi\left| 0 \right \rangle$. 
The $4$ point correlation function of $\phi$ on $\Sigma_2$ is given by 
\begin{equation}
\begin{split}\label{n2con}
&\left\langle \phi(r_1, \theta_1)\phi(r_2, \theta_2)\phi(r_1, \theta_1+2\pi)\phi(r_2, \theta_2+2\pi)\right \rangle_{\Sigma_2^{con}} \\
&=\left\langle \phi(r_1, \theta_1)\phi(r_2, \theta_2)\phi(r_1, \theta_1+2\pi)\phi(r_2, \theta_2+2\pi)\right \rangle_{\Sigma_2}+
\sum_{k=1}^{\infty}\left \langle \mathcal{O}^{2n} :\frac{\left( c_2 \int_{\partial A} d{\bf x}^{d-1}\phi^2(0,0, {\bf x})\right)^k}{k!}:\right \rangle_{\Sigma_n}.
\\
\end{split}
\end{equation}
The correlation function between the origin and the locations of operators are given by
\begin{equation}
\left\langle\phi(r_1,\theta_1,{\bf x}_1)\phi(0, \theta_2, {\bf x}_2)\right \rangle =\frac{1}{4\pi~^2\left(r_1^2 +\left|{\bf x}_1-{\bf x}_2\right|^2\right)}.
\end{equation}
The contribution from this correlation function is at most given by 
\begin{equation}
\left\langle\phi(r_1,\theta_1,{\bf x}_1)\phi(0, \theta_2, {\bf x}_2)\right \rangle \sim \frac{1}{4\pi^2 r_1^2}.
\end{equation}

After performing the analytic continuation, the contribution from this propagator is at most $\mathcal{O}(1)$ in $\epsilon \rightarrow 0$ limit and $\int_{\partial A} d{\bf x}^{d-1}$ is regularized. Therefore the leading term of $\left\langle \phi^4\right \rangle_{\Sigma_2}$ is $\mathcal{O}\left(\epsilon^{-4}\right)$ and the second term in (\ref{n2con}) is at most $\mathcal{O}(\epsilon^{-2})$ in this limit. 
Then it can not contribute to the 4 point correlation function of $\phi$. Therefore we can ignore the effect of the conical singularity.

\section{Discussions and Conclusions}
In the present paper, we studied the (R$\acute{e}$nyi) entanglement entropy for a new class of excited states. These excited states are defined by acting various local operators on the ground state. Therefore we can investigate the evolution of (R$\acute{e}$nyi) entanglement entropies with time.

Furthermore we defined the excess of (R$\acute{e}$nyi) entanglement entropies for those locally excited states.
We chose a half $(x^1 \ge 0)$ of the total space as the subsystem A in present case.
Those entropies are related to the correlation functions of the operators.
In general, it is difficult to compute the (R$\acute{e}$nyi) entanglement entropy in interacting filed theories. 
For this excess of (R$\acute{e}$nyi) entanglement entropies, all that we have to do is to calculate the correlation functions of the operators if we compute them perturbatively.
It is expected that this will be perturbatively computable in interacting filed theory.

By the replica method we computed (R$\acute{e}$nyi) entanglement entropies for excited states defined by acting particular operators on the ground state in free massless scalar field theory in even dimension. 
We performed an analytic continuation to real time and investigated their time evolution.
The (R$\acute{e}$nyi) entanglement entropies for locally excited states approach some finite constants at late time\footnote[11]{In local quenches, entanglement entropy keeps to increase at late time and never approaches to a particular constant.}. We found their time evolution can be interpreted in terms of relativistic propagation of the entangled pair. 

We gave an explanation of the final value of the (R$\acute{e}$nyi) entanglement entropies for those locally excited states in terms of entangled pair interpretation. We defined the (R$\acute{e}$nyi) entanglement entropies of operators by the final value of the (R$\acute{e}$nyi) entanglement entropies for those locally excited states. They characterize operators in the view point of quantum entanglement and they are independent of the conformal dimension of operators. The (R$\acute{e}$nyi) entanglement entropies of $:\phi^k:$ are given by (\ref{gfo}). They do not depend on the spacetime dimension. They do not change if we deform the shape of subsystem A continuously. They can be defined in any dimensions.

We obtained various new results that we summarize below.
In previous paper \cite{pne}, we computed the n-th R$\acute{e}$nyi entanglement entropies for the states which are excited by $:\phi^k:$ by replica method when $n, k$ is small.
In this paper, we obtained the explanation which shows that the (R$\acute{e}$nyi) entanglement entropies of$:\phi^k:$ are given by those of binomial distribution by the replica trick for any $n$ and $k$ and any dimensions. We checked that they agree with the results which we obtained in terms of entangled pair.
We argued the (R$\acute{e}$nyi) entanglement entropies of particular operators are also given by those of binomial distribution. These operators are composed of single-species operators such as $:(\partial^m \phi)^n:$ and we call them S-operators. We studied the (R$\acute{e}$nyi) entanglement entropies of $:\phi^k:$ in large $k$ limit.

We also obtained the sum rule for the final value of the (R$\acute{e}$nyi) entanglement entropies for locally excited states in any dimensions. Here those locally excited states were defined by acting various local operators on the ground state. We assumed these operators were separated spatially and they were given by $:\phi^k:$. We checked that the sum rule which we obtained by the replica method agrees with the one derived from the entangled pair. We argued that $\Delta S^{(n) f}_A $ for the states defined by acting general operators on the ground state obey the same sum rule in any dimension. Here we assumed that those operators were located separately.

\begin{table}[b] 
  \caption{The second (R$\acute{e}$nyi) entanglement entropy of $:\phi\partial_r\phi :$ in the $\epsilon \rightarrow 0$ limit.}
  \begin{tabular}{|c|c|c|c|} \hline
& Dimension of Spacetime &Entanglement of Operators \\ \hline \hline
                                     & 4 & $ \log{\left[\frac{100}{37}\right]} $ \\ \cline{2-3}
Second R$\acute{e}$nyi Entropy      & 6& $\log{\left[\frac{324}{121}\right]}$  \\ \cline{2-3}
                                     & 8 &  $ \log{\left[\frac{676}{253}\right]}$ \\ \cline{2-3} 
\hline  \end{tabular}
\end{table}

We would also like to list a few of open problems.
As we mentioned above, we found the evidence that the (R$\acute{e}$nyi) entanglement entropies of S-operators will be given by (\ref{gfo}).
Let us compute the (R$\acute{e}$nyi) entanglement entropies of the operators which are composed of multi-species operators  such as $:\phi \partial \phi:$, $:\phi \partial \partial \phi:$ and so on. They are not S-operators.
We call them M-operators.
It is expected that the (R$\acute{e}$nyi) entanglement entropies of M-operators will not obey the formula given in (\ref{gfo}).
For example, we prepare the state defined by acting an M-operator $:\phi \partial_r \phi:$ on the ground state,
\begin{equation} 
\left| \Psi \right \rangle =\mathcal{N}^{-1} :\phi \partial_r \phi :\left|0 \right \rangle.
\end{equation}
The (R$\acute{e}$nyi) entanglement entropy of $:\phi \partial_r \phi:$ depends on the spacetime dimension as in Table.2.
The time evolution of $\Delta S^{(2)}_A$ for this state is plotted in Fig.15.
The second R$\acute{e}$nyi entanglement entropy does not obey the formula in (\ref{gfo}).

Let us interpret it in terms of entangled pair.  We can decompose operators to the left moving modes and right moving modes respectively,
\begin{equation}
\begin{split}
&\phi =\phi_L+\phi_R, \\
&\partial_r \phi =\partial_r \phi_L +\partial_r \phi_R. \\
\end{split}
\end{equation}
In this case, the relation between $\phi$, $\partial_r \phi$ is not clear.
Then these operators are probably not independent of each other.
As we explained, the second R$\acute{e}$nyi entanglement entropy of $:\phi \partial_r \phi:$ depends on the spacetime dimension and does not obey the formula in (\ref{gfo}).
Therefore, it is interesting to understand why $\Delta S^{(2)}_A$ depends on spacetime dimensions and to interpret it physically.
It is one of the open problems.
 
\begin{figure}\label{l10dep}	
  \centering
  \includegraphics[width=8cm]{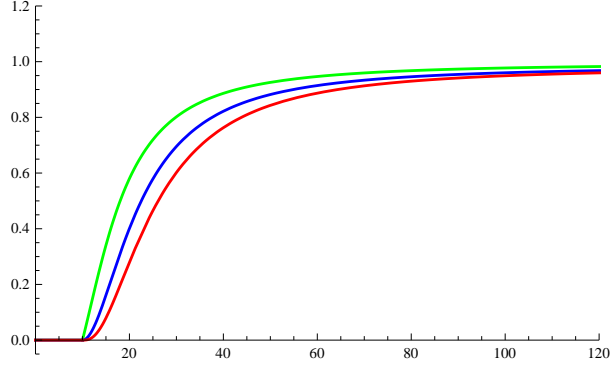}
  \caption{The evolution of $\Delta S^{(2)}_A$ with $t_1$ in the $\epsilon \rightarrow 0$ limit.
The vertical axis corresponds to $\Delta S^{(2)}_A$. The horizontal axis corresponds to $t_1$. Here we chose $l=10$.
The green curve is the time evolution of $\Delta S^{(2)}_A$ for $\phi\partial_r \phi\left|0\right \rangle$ in $4$ dimension.
The blue curve is the time evolution of $\Delta S^{(2)}_A$ for that state in $6$ dimension. The red curve is the time evolution of $\Delta S^{(2)}_A$ for that state in $8$ dimension.
}
\end{figure}

Other interesting open problems include:
\begin{itemize}
\item[-] It is interesting to find the formula for the (R$\acute{e}$nyi) entanglement entropies of M-operators

\item[-] It is interesting to investigate the time evolution of (R$\acute{e}$nyi) entanglement  entropy for the locally excited states in interacting field theories.

\item[-] It will be also interesting to investigate the (R$\acute{e}$nyi) entanglement entropies of operators in massive theory.
\end{itemize}

Also we would like to refer to \cite{paw} for large N CFT and refer to \cite{ttq} for rational CFTs in two dimension.

\section*{Acknowledgements}
MN would like to thank  the collaborators T. Takayanagi, and T. Numasawa for reading the draft of this paper and giving us very useful and remarkable comments. MN also thanks P. Caputa, S. He and N. Shiba for reading the draft of this paper and giving us useful and remarkable comments.
MN is supported by JSPS fellowship.

\end{document}